\documentclass[fleqn,usenatbib]{mnras}
\usepackage{pdflscape}

\usepackage{newtxtext,newtxmath}

\usepackage[T1]{fontenc}

\DeclareRobustCommand{\VAN}[3]{#2}
\let\VANthebibliography\thebibliography
\def\thebibliography{\DeclareRobustCommand{\VAN}[3]{##3}\VANthebibliography}
\usepackage{graphicx}	
\usepackage{amsmath}	

\usepackage{xcolor}

\usepackage[normalem]{ulem}
\newcommand{\hl}{\textcolor{black}}

\title[SFH of the SMC's northeastern shell]{\textcolor{black}{Unveiling the purely young star formation history of the SMC's northeastern shell from colour-magnitude diagram fitting}}

\author[Sakowska et al.]{Joanna~D.~Sakowska$^{1}$\thanks{E-mail: \url{j.sakowska@surrey.ac.uk}},
Noelia~E.~D.~No\"el$^{1}$,
Tom\'as~Ruiz-Lara$^{2, 3}$, 
Carme~Gallart$^{4, 5}$,
Pol~Massana$^{6}$,
\newauthor
David~L.~Nidever$^{6, 7}$,
Santi~Cassisi$^{8, 9}$,
Patricio~Correa-Amaro$^{10}$,
Yumi~Choi$^{7}$, 
Gurtina~Besla$^{11}$,
\newauthor
Denis~Erkal$^{1}$,
David~Mart\'inez-Delgado$^{12}$,
Matteo~Monelli$^{4, 5, 13}$,
Knut~A.~G.~Olsen$^{7}$,
Guy~S.~Stringfellow$^{14}$
\newauthor
and the SMASH collaboration
\\ 
$^{1}$Department of Physics, University of Surrey, Guildford GU2 7XH, UK\\
$^{2}$Universidad de Granada, Departamento de Física Teórica y del Cosmos, Campus Fuente Nueva, Edificio Mecenas, E-18071,
Granada, Spain\\
$^{3}$Instituto Carlos I de F\'isica Te\'orica y Computacional, Universidad
de Granada, E-18071 Granada, Spain\\
$^{4}$IAC- Instituto de Astrof\'isica de Canarias, Calle V\'ia Lactea s/n, E-38205 La Laguna, Tenerife, Spain\\
$^{5}$Departmento de Astrof\'isica, Universidad de La Laguna, E-38206 La Laguna, Tenerife, Spain\\
$^{6}$Department of Physics, Montana State University, P.O. Box 173840, Bozeman, MT 59717-3840\\
$^{7}$NSF's National Optical-Infrared Astronomy Research Laboratory, 950 North Cherry Ave, Tucson, AZ 85719\\
$^{8}$INAF – Osservatorio Astronomico di Abruzzo, Via M. Maggini, 64100 Teramo, Italy\\
$^{9}$INFN - Sezione di Pisa, Universitá di Pisa, Largo Pontecorvo 3, 56127 Pisa, Italy\\
$^{10}$Institute for Astronomy, University of Edinburgh, Royal Observatory, Blackford Hill, Edinburgh EH9 3HJ, UK\\
$^{11}$Steward Observatory, University of Arizona, 933 North Cherry Avenue, Tucson AZ, 85721\\
$^{12}$Instituto de Astrof\'isica de Andaluc\'ia, CSIC, E-18080, Granada, Spain\\
$^{13}$INAF - Osservatorio Astronomico di Roma, via Frascati 33, 00078 Monte Porzio Catone, Italy\\
$^{14}$Center for Astrophysics and Space Astronomy, University of Colorado, 389 UCB, Boulder, CO 80309-0389, USA}
\date{Accepted 2024 July 18. Received 2024 June 29; in original form 2024 January 22}

\pubyear{2024}

\begin{document}
\label{firstpage}
\pagerange{\pageref{firstpage}--\pageref{lastpage}}
\maketitle

\begin{abstract}
\textcolor{black}{We obtain a quantitative star formation history (SFH) of a shell-like structure (`shell') located in the northeastern part of the Small Magellanic Cloud (SMC). We use the Survey of the MAgellanic Stellar History (SMASH) to derive colour-magnitude diagrams (CMDs), reaching below the oldest main-sequence turnoff, from which we compute the SFHs with CMD fitting techniques. We present, for the first time, a novel technique that uses red clump (RC) stars from the CMDs to assess and account for the SMC’s line-of-sight depth effect present during the SFH derivation. We find that accounting for this effect recovers a more accurate SFH. We quantify a $\sim$7 kpc line-of-sight depth present in the CMDs, in good agreement with depth estimates from RC stars in the northeastern SMC. By isolating the stellar content of the northeastern shell and incorporating the line-of-sight depth into our calculations, we obtain an unprecedentedly detailed SFH. We find that the northeastern shell is primarily composed of stars younger than $\sim$500 Myrs, with significant star formation enhancements around $\sim$250 Myr and $\sim$450 Myr. These young stars are the main contributors to the shell's structure. We show synchronicity between the northeastern shell's SFH with the Large Magellanic Cloud's (LMC) northern arm, which we attribute to the interaction history of the SMC with the LMC and the Milky Way (MW) over the past $\sim$500 Myr. Our results highlight the complex interplay of ram pressure stripping and the influence of the MW's circumgalactic medium in shaping the SMC's northeastern shell.}

\end{abstract}


\section{Introduction}
\label{sec:intro}

The faint peripheries of galaxies contain stellar fossil records of their mass assembly history (e.g., through galactic mergers, accretions and/or dynamical interactions with other galaxies) and, hence, harbour important clues to understanding the galaxy's formation and evolution (e.g., \citealt{ElmegreenHunter2017}). As such, star formation history (SFH) studies of galactic peripheries were we can resolve the individual stars can provide important information on the processes that govern galactic growth. \hl{In particular, by reaching the oldest main-sequence turnoffs (oMSTO) in colour-magnitude diagrams (CMDs) of resolved stellar populations, CMD fitting techniques can break the age-metallicity degeneracy leading to accurate determination of their SFHs (e.g., \citealt{Gallart2005}).}

The outskirts of the Small and Large Magellanic Clouds (SMC and LMC), located at $\sim$60 kpc and $\sim$50 kpc away from us (\citealt{Pietrzynski2019, Graczyk2020}), offer an outstanding opportunity to study SFHs in exquisite detail as we can obtain precise photometry of individual stars in these galaxies down to the oMSTO using ground-based telescopes \hl{(e.g., STEP: \citealt{Ripepi2014};  SMASH: \citealt{nidever2017}; VISCACHA: \citealt{Maia2019}; YMCA: \citealt{Gatto2020}; DELVE: \citealt{drlicawagner2021})}. 
The interacting history of the Magellanic Clouds (MCs) makes this system even more beguiling. For instance, such interplay between the MCs led to the emergence of \hl{large-scale morphological features such as} the Magellanic Bridge (\citealt{HindmanKerrMcGee1963, Noel2013}), the Leading Arm (\citealt{Putman1998}), and the Magellanic Stream (\citealt{Mathewson1974}).  
\\
\hl{Given the SMC’s smaller total mass (\citealt{DeLeo2023}) it is plausible that the LMC’s gravitational influence played a key role in tidally shaping the SMC (e.g. \citealt{deleo2020}).} Indeed, the SMC's outskirts are home to a deluge of stellar structures such as the young ($\sim$150 Myr) northeastern shell-like overdensity (`northeastern shell', \citealt{martinezdelgado2019}, hereafter MD19), the SMCNOD overdensity $\sim8^{\circ}$ northwest of the SMC (\citealt{Pieres2017}), the various potential stellar streams in the northwest outskirts of the MCs (\citealt{Belokurov2016, Navarrete2019}), \hl{stellar evidence for a tidal counterpart of the Magellanic Bridge (Counter-bridge: \citealt{Dias2021}), and a structure in the western outskirts confirmed to be moving away from the SMC (West Halo: \citealt{Dias2016}; \citealt{Niederhofer2018}; \citealt{Zivick2018}; \citealt{Piatti2021}; \citealt{Dias2022}). Hence, the complex peripheries of the SMC merit a thorough study of their SFHs to help us elucidate key elements of its formation and evolution.} In particular, the coherent SMC's northeastern shell overdensity, already noted in photographic plates from the 1950s \citep[see][]{Devaucouleurs1972}, and subsequently confirmed by others (\citealt{Brueck1978}; \citealt{Albers1987}) has been the subject of studies since its detection.
\\ 
Analysing shallow CMDs, \citet[][]{Brueck1978} hinted at the presence of a young population in this region (called ``outer-arm'' in their work). Using deeper CMDs from the SMASH survey, MD19 showed that the northeastern shell stands out when the spatial distribution density map of younger populations (upper main-sequence stars) is depicted rather than when the \hl{intermediate-age and} older populations are presented (see Fig.~\ref{fig:FigRegions} below). \hl{To disentangle the nature of the northeastern shell, MD19 applied the colour function method (\citealt{Noelia2007}) and found hints of not only young but also intermediate-age ($\sim$1.5 Gyr - 6 Gyr) and old ($\sim$8 Gyr - 13.5 Gyr) stellar populations. Analysis of control regions near the northeastern shell highlighted its young stellar populations to be in stark contrast to the control regions. Guided by the ages of the Classical Cepheids (CCs) and young star clusters within the northeastern shell, MD19 suggested that the structure formed in a recent SF episode likely triggered by an interaction between the MCs around $\sim$150 Myr ago \citep{Choi2022}. Using the colour function method, MD19 were not able to date the intermediate-age and old populations further and suggested that these populations could be contamination from the SMC's field stars.}
\hl{\cite{Piatti2022} performed a quantitative analysis of the star clusters on the northeastern shell (and surrounding close-by regions) using the SMASH survey by constructing their CMDs, cleaned from field stars, and employing CMD fitting techniques. The authors dated the star clusters to be as young as $\sim$30 Myrs old, evidencing very recent star formation in the region. Despite the careful quantitative analysis, the authors were naturally limited by the number of star clusters available in comparison to field stars. \cite{Hota2024} cross-matched far-ultraviolet (FUV) stars with optical Gaia EDR3 data (\citealt{gaia2021}) and performed a qualitative analysis by visually overlapping isochrones on the FUV-optical CMDs, identifying $\sim$60 Myr and $\sim$260 Myr old enhancements. However, the FUV stars selected were younger than 400 Myrs and therefore the CMDs did not reflect all of the stellar populations present within the northeastern shell.
To comprehensively examine the northeastern shell’s stellar content, accurately date the young, intermediate-age, and old populations, and shed further light on its origin, a quantitative SFH determination of all of the stars within the northeastern shell is required.}
\\
Obtaining accurate SFHs for the SMC's peripheral region is challenging mainly due to the known line-of-sight depths variations in this galaxy (see, e.g., 
\citealt{HatziHawkins1989},
\citealt{Hatzidimitriou1989}, \citealt{Gardiner1991}, 
\citealt{Gardiner1992}, \citealt{Crowl2001}). Such line-of-sight depths create an extra layer of observational effects on the observed CMDs that must be taken into account in quantitative SFHs determinations. Before obtaining the SFH we must then assess the depth in the line-of-sight of the region of interest. Tracing stellar populations of different ages, variable stars constitute outstanding objects to accurately measure the line-of-sight depths across the SMC (e.g., \citealt{Hatzidimitriou1989}). 
Mapping the young stellar population ($\sim$500 Myr) of the SMC, CCs illustrate that this galaxy is tilted and elongated \hl{with its eastern side $\sim$20 kpc closer to the LMC than its western part (\citealt{Scowcroft2016}; \citealt{Ripepi2017}; \citealt{Jacyszyn-Dobrzeniecka2016}). }
RR Lyrae -tracers of old stellar populations ($>$ 10 Gyr)- are statistically more numerous than CCs in the SMC ($\sim$22,859 RR Lyrae versus $\sim4663$ CCs in the SMC OGLE-IV catalogues, \citealt{Jacyszyn-Dobrzeniecka2016, Jacyszyn-Dobrzeniecka2017}) and show an ellipsoidal distribution along the SMC's line-of-sight (\citealt{Haschke2012,SubramanianSubramaniam2012, Deb2015}; \citealt{Deb2015, Jacyszyn-Dobrzeniecka2017}) resulting in measurements ranging from $\sim$1 kpc to $\sim$10 kpc (\citealt{Muraveva2018}). 
\\
While variable stars have been pivotal in mapping the complex structure of the SMC, the stellar populations of the northeastern shell contain a broad range of ages (not represented by CCs and RRLs) forcing us to find alternative indicators to estimate the line-of-sight depth with. An optimum alternative is to use the magnitude spread of the red clump (RC) stars in the SMC's CMD. Given that the RC has a very narrow magnitude range \citep{Salaris2005}, its shape is the part most conspicuously affected by line-of-sight depths on CMDs, causing it to appear vertically `smudged' in magnitude. The RC in observed CMDs has been previously used to obtain SMC's line-of-sight depths (e.g. \citealt{Hatzidimitriou1989}) in spite of the fact that the magnitudes of the RC stars are affected by distance, photometric errors, age and metallicity (e.g., \citealt{Ata1999AJ}; \citealt{Girardi2001}). With the advent of more precise photometric surveys covering wider areas across the SMC (\citealt{nidever2013}; \citealt{Tatton2021}; \citealt{ElYoussoufi2021}) the RC became a prime alternative to assert line-of-sight variations across the SMC. 
\\
\hl{We present here the first quantitative SFH determination of the SMC's northeastern shell, as derived from its field stars, with CMD fitting techniques.} To achieve our goals, we use data from the second and final release of the Survey of the MAgellanic Stellar History \citep[SMASH;][]{nidever2017} and \hl{introduce, for the first time, the line-of-sight depth effect during the SFH derivation.}
\\
This paper is organised as follows. In Section \ref{sec:data} we present the SMASH data and the selection of the different spatial regions.  The SFH determination methods, including the consideration of the line-of-sight depth effects, are described in Section \ref{sec:sfhprocedure}. In Section \ref{sec:sfh} we show our SFH results and in Section \ref{sec:discussion} we discuss the implications. In Section \ref{sec:conclusions} we draw the main conclusions. Finally, we support our methodology with Appendix \ref{sec:sfhmocksappendix} where we show the various SFHs recovery tests, including different variations in the line-of-sight depth.

\section{DATA}
\label{sec:data}
\subsection{The SMC's northeastern shell in SMASH}

The Survey of the MAgellanic Stellar History (SMASH; \citealt{nidever2017}) used the Dark Energy Camera \citep[DECam;][]{Flaugher2015} installed on the Blanco 4-m telescope at the CTIO (Cerro Tololo Inter-American Observatory) in Chile. Thanks to DECam's large ($\sim$3 deg$^2$) field of view, SMASH surveyed across $\sim 2400 \ \mathrm{deg}^2$ of the Magellanic System, resulting in a net $\sim 480 \ \mathrm{deg}^2$ of high-quality photometric data of the SMC. The contiguous, deep (\textit{ugriz} $\sim$ 24.5 mag) optical coverage of the MCs' main bodies and peripheries achieved by SMASH are optimum to study their stellar structures and SFHs. 

We use the second and final SMASH data release (SMASH DR2; \citealt{nidever2021}) to study the SMC's northeastern shell reported in MD19, located across \hl{SMASH Field 14 ($\alpha=$01:20:26.78, $\delta=$-71:15:32.76)}. In Section \ref{subsec:regionselection} we describe in detail how we spatially select the northeastern shell \hl{and a control region surrounding it}. We refer the reader to Figure \ref{fig:FigRegions} for a stellar density map highlighting the studied regions (see also Figure 1 of \citealt{nidever2021} as a reference for SMASH fields).

The SMASH DR2 catalogue has several columns generated by PHOTRED (see \citealt{nidever2017} for details on SMASH image reduction) which are used to constrain our photometric selection in the $g$ and $i$ bands. We set a $-2.5 < \mathrm{SHARP} < 2.5$ constraint to reduce the contamination by galaxies and spurious objects and the photometric uncertainties in the $g$ and $i$ bands were limited to $0.3$ mag (as tested and proven effective in \citealt{ruizlara2020, Massana2022} for SFH studies using data from SMASH). Dust correction was applied using a reddening map constructed with RC stars following the techniques described in \cite{Choi2018a}, assuming an intrinsic $g-i$ colour of 0.72. We adopted a distance modulus to the SMC of $(m - M_{o}) = 18.96$ \citep{deGrijs2015}.

\subsection{Artificial star tests}
\label{subsec:asts}

In order to evaluate the photometric errors and completeness of our data, caused by stellar crowding, blending, and other measurement errors, we performed artificial star tests (ASTs). The ASTs consist of injecting stars with known magnitudes into the DECam images and re-calculating the photometry using \texttt{PHOTRED} \cite{nidever2017}, as done for the observed data. The resulting magnitudes are then compared with their initial values, quantifying the observed photometric errors and the completeness of our observations (see Sect.~\ref{subsec:sfhsolve}). \hl{We use the observed distribution of stars in each of the survey fields to create artificial star catalogues containing stars covering a wide range of colours, magnitudes and on-sky distributions.} The procedure applied in the SMASH data for computing ASTs is discussed in more detail in \cite{Monelli2010a} and \cite{Rusakov2021}.
We injected $\sim 2.1 \times 10^{6}$ stars in {\hl{SMASH Field 14 \citep{nidever2021}}} and calculated the completeness in each region based on the results of the ASTs. The results can be observed in Figure~\ref{fig:ASTs}: the completeness \hl{in Region A} is 90\% at $g\sim$ 3.6, $i\sim$ 3.1 and 50\% at $g\sim$ 5.65, $i\sim$ 4.9. \hl{Stars near the oMSTO, located at $i\sim$3.0, are present in the CMD with 91\% completeness. This excellent completeness at the oMSTO puts us in a prime position to derive SFHs.}

\subsection{Region selection}
\label{subsec:regionselection}

\begin{figure}
\includegraphics[scale=0.46]{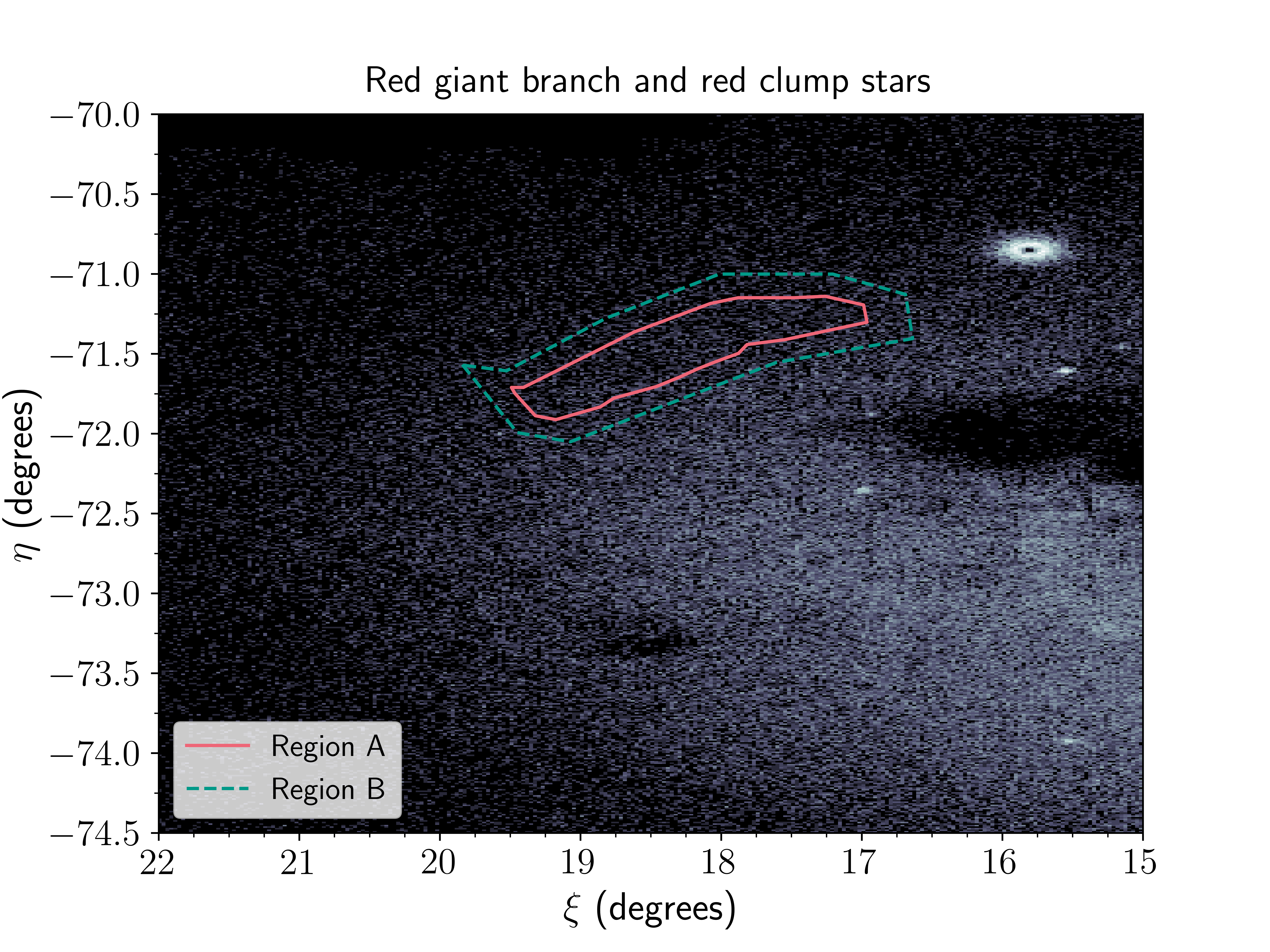}
\includegraphics[scale=0.46]{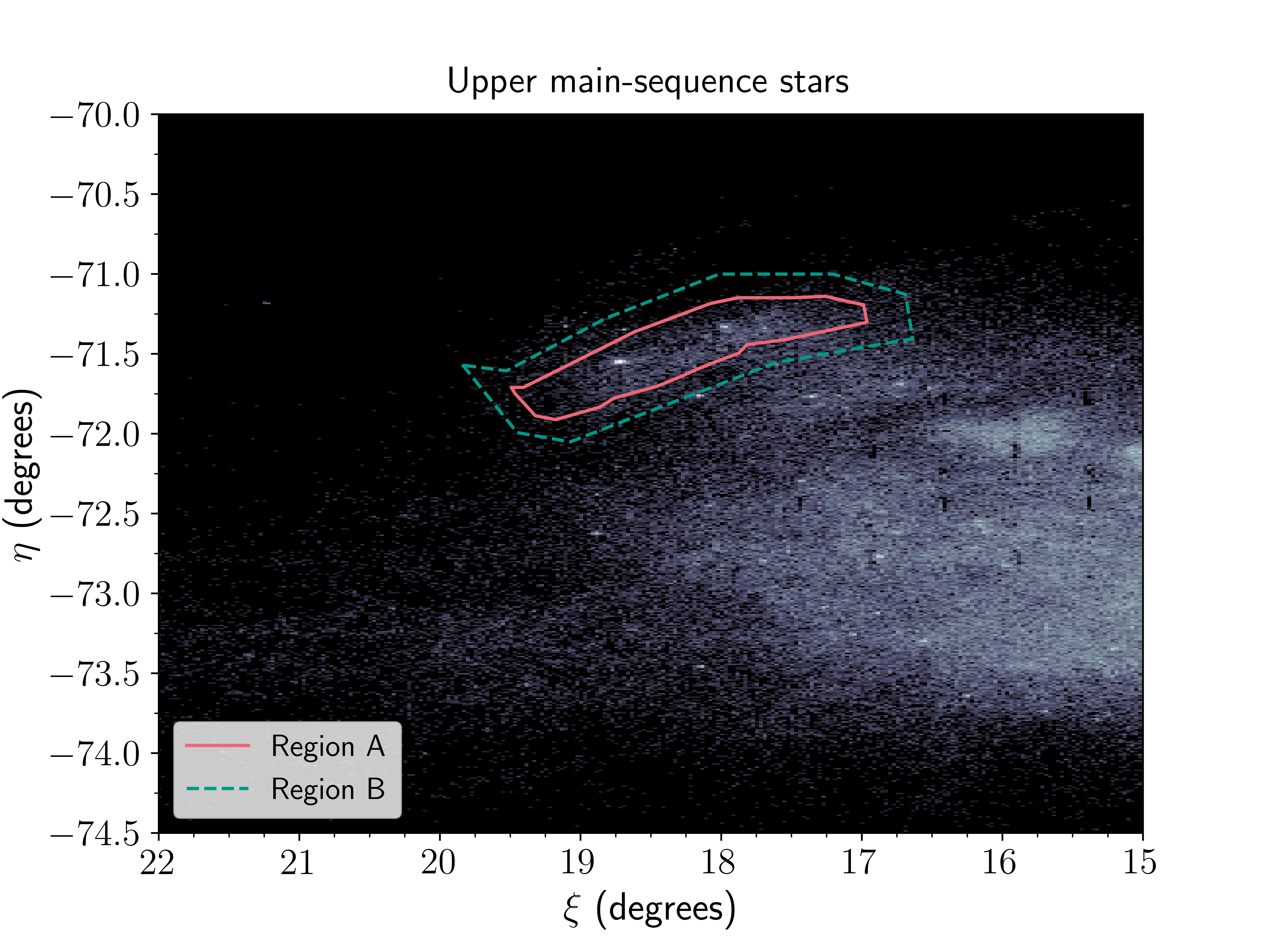}
\caption{Top: Spatial distribution of selected \hl{intermediate-age and} older populations (red giant branch and red clump stars) across the SMC with $\sim$ 655,558 stars. Bottom: density map of the younger populations (upper main-sequence stars) with $\sim$ 324,535 stars. \hl{The region containing the northeastern shell (region A) is outlined in pink, and the region used to estimate the contribution of SMC field stars to Region A is in a dashed green line (region B).}}\label{fig:FigRegions}
\end{figure}

\begin{figure}
\begin{center}
\includegraphics[width=0.49\textwidth]{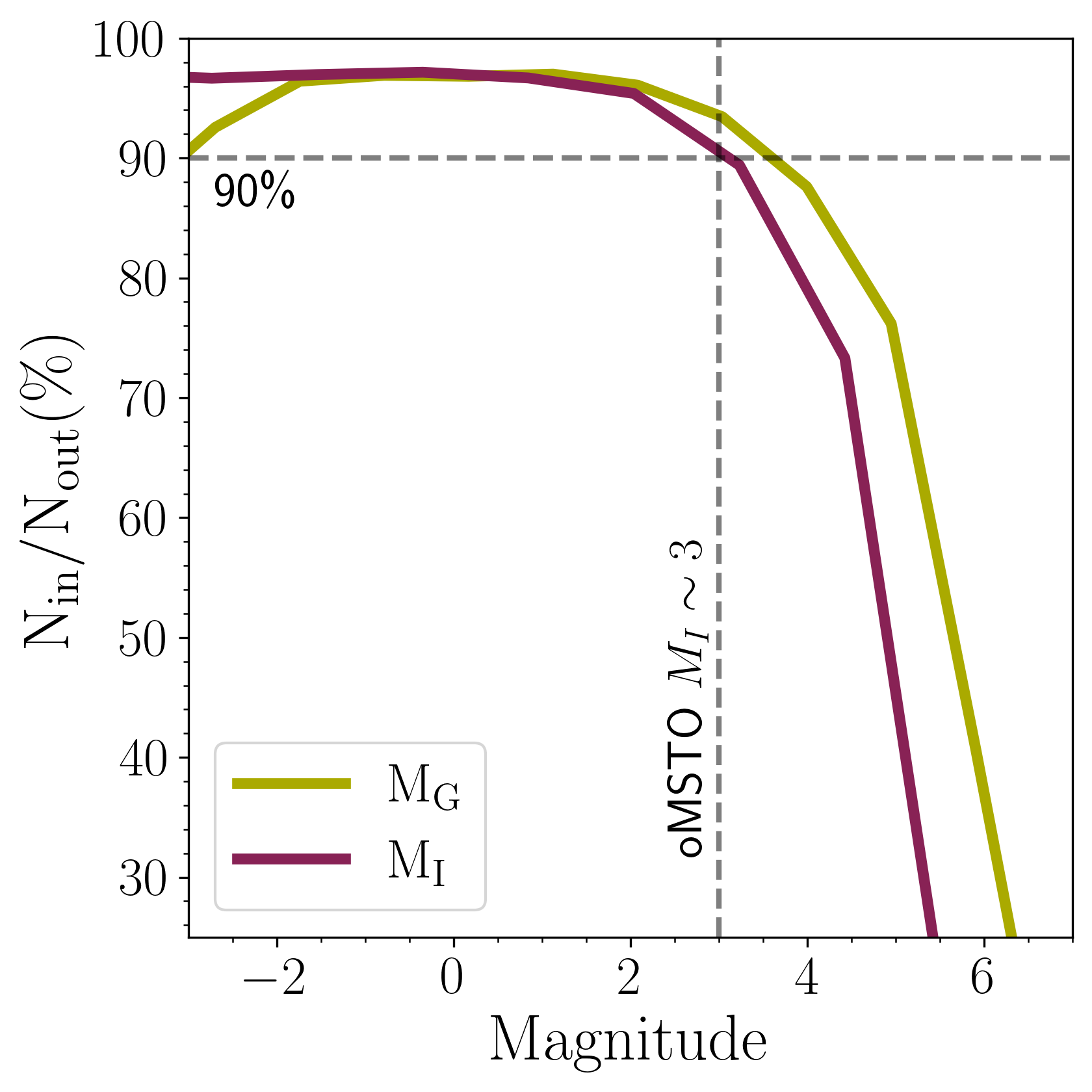}
\caption{\hl{The results of our artificial star tests for region A in the \textit{g, i} bands. $N_{in}/N_{out} (\%)$ denotes the ratio of injected versus recovered stars of known magnitude as a percentage. We mark the 90$\%$ completeness threshold (black dashed lines). We also denote the magnitude threshold of oMSTO stars ($M_{i}\sim$3.0), showing our excellent $>$90$\%$ completeness.}}
\label{fig:ASTs}
\end{center}
\end{figure}

\begin{figure}
\begin{center}
\includegraphics[scale=0.45]{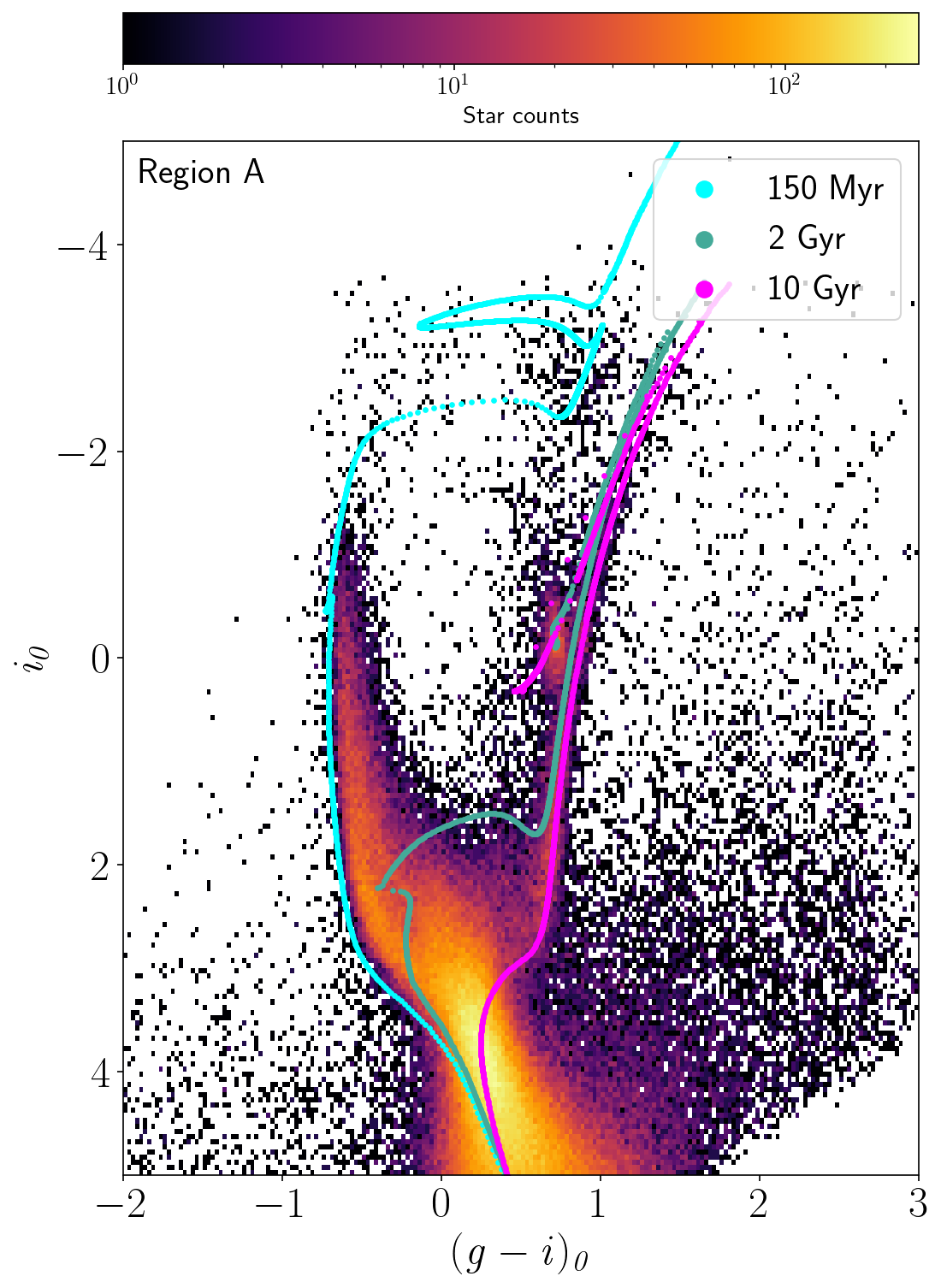}
\caption{CMD corresponding to \hl{region A}  (see Figure \ref{fig:FigRegions}). We overlaid isochrones from the BaSTI-IAC library with three different ages and metallicities: a young isochrone shown in cyan (Z $=$ 0.002, age $=$ 150 Myr), an intermediate-age isochrone presented in green (Z $=$ 0.002, age $=$ 2 Gyr), and an old isochrone depicted in magenta (Z $=$ 0.001, age $=$ 10 Gyr).}
\label{fig:FigCMDs}
\end{center}
\end{figure} 

To visualise the \hl{northeastern shell} we followed the procedure presented in MD19. To assess how the stars are spatially distributed, we isolated SMC stars located in different regions of the combined $i$ and $g-i$ CMDs of SMASH \hl{Field 14 and its neighbouring SMASH Fields 9 ($\alpha= $01:01:27.40, $\delta= $-70:43:05.51), 10 ($\alpha=$01:03:36.32, $\delta=$-72:18:54.4), and 15 ($\alpha=$01:24:33.52, $\delta=$-72:49:30.00)}. 
In the top panel of Figure~\ref{fig:FigRegions}, we present the red giant branch (RGB) and RC stars of the northeastern SMC (indicators of intermediate-age and old populations). In the bottom panel of Figure~\ref{fig:FigRegions} we show stars occupying the upper young main-sequence (MS) region (of the same fields SMASH 14, 9, 10, and 15), noting that the \hl{northeastern shell} is only discernible when isolating these young MS stars. 

\hl{To isolate the northeastern shell's stellar content we divided the area into two regions as seen in Fig.~\ref{fig:FigRegions} (see figure caption for more details). The region marked with the solid pink polygon corresponds to region `A' that contains both the SMC's northeastern shell and stellar contamination from the SMC's main body. Given that we are viewing the northeastern shell `face on', it is impossible to visually disentangle which stars belong to the northeastern shell and which stars belong to the SMC's main body. As such, we select a region around the northeastern shell -region `B'- depicted with a green dashed polygon. Region B represents the stellar population of the SMC's field stars present around the 
northeastern shell (an annulus-like region). By obtaining the SFH of region B, we obtain the best approximation possible for the SFH contribution from the SMC's field stars. So subtracting the SFH of region B from that of region A results in a final, `clean' SFH of the northeastern shell, with as little contamination from the SMC's field stars as possible.}

\hl{In Figure \ref{fig:FigCMDs} we present the observed SMASH CMD corresponding to region A} in Figure \ref{fig:FigRegions}. 
To visually illustrate the stellar populations present, we overlay young (Z = 0.002, age = 150 Myr), intermediate-age (Z = 0.002, age = 2 Gyr), and  old (Z = 0.001, age = 10 Gyr) isochrones from the \verb|BaSTI-IAC| library\footnote{\url{http://basti-iac.oa-abruzzo.inaf.it}} (\citealt{Pietrinferni2024}). We adopted the metallicities from MD19 as they are consistent with spectroscopic determinations using young CCs and HII regions (\citealt{Russell1992, Romaniello2009, Lemasle2017}). The selected stellar model predictions correspond to the model set accounting for the solar-scaled heavy element mixture, convective core overshooting, efficient atomic diffusion and mass loss efficiency fixed at $\eta$ = 0.3 (see \citealt{Hidalgo2018} for more details). 
\hl{The CMD of region A is} very well populated by stars of all ages (including young, intermediate-age and old stars) and metallicities. \hl{The overlaid isochrones highlight the young nature of the northeastern shell but also the presence of intermediate-age and old populations in the same field of view.}

\section{Star Formation History procedure}
\label{sec:sfhprocedure}

In this Section we describe in detail a novel approach to compute SFHs taking into account line-of-sight depth effects using existing codes. This approach consists of a two-step SFH recovery. In a first step, we solve for a representative SFH of the region under study without simulating line-of-sight depth effects. In the second step, we assess the line-of-sight depth using the comparison of observed and simulated RC (taking into account approximate age and metallicity distributions), and obtain the final SFH considering the line-of-sight depth.

\subsection{Solving for the star formation history: Standard procedure (step 1)}
\label{subsec:sfhsolve}

We created individual synthetic CMDs for a robust comparison between the observed CMDs and theoretical models. To construct the synthetic CMDs \hl{for regions A and B} we used the solar-scaled \verb|BaSTI-IAC| stellar evolution models (\citealt{Pietrinferni2021}), and generated a global synthetic population containing $5 \times 10^{7}$ stars with a flat distribution at birth in age and metallicity ranging from $0.03$ to $14$ Gyr in age and $0.00001$ to $0.025$ in metallicity (Z). Following \cite{ruizlara2020} and \cite{Massana2022}, we assumed a Kroupa initial mass function (IMF; \citealt{Kroupa2001}) and a binary fraction of 50$\%$ with a mass ratio ranging from 0.1 to 1.  
We simulated observational effects on these synthetic CMDs, modelled using the \verb|DisPar| code. \verb|DisPar| utilises the ASTs results relevant to the region at hand to `disperse' the stars from their actual positions on the synthetic CMDs, according to the measured observational errors and completeness (see Appendix B of \citealt{ruizlara2021} for an application and detailed description of \verb|DisPar|). To obtain the best solution of the SFH per region studied we used the \verb|THESTORM| code (\citealt{Bernard2015b, Bernard2018}). \verb|THESTORM| uses a Poisson adapted $\chi^{2}$ \citep{Cash1979} to find the best combination of simple stellar populations (SSP) from the dispersed synthetic CMD that fits the distribution of stars in the observed CMD. The set of SSPs that we have used (396 SSPs in total) uses the following age-metallicity grid:

\begin{itemize}
    \item Age: [0.03 - 0.1; 0.1 - 1.0 in steps of 0.1; 1.0 - 2.0 in steps of 0.2; 2.0 - 2.6 in steps of 0.3; 2.6 - 3.0; 3.0 - 10.0 in steps of 0.5; 10.0 - 13.0 in steps of 1; 13.0 - 13.9] Gyr
    \item Z: [0.1, 1, 6, 16, 30, 45, 65, 90, 120, 160, 200, 249] $\times10^{- 4}$
\end{itemize}

\begin{figure*}
\begin{center}
\includegraphics[scale=0.52]{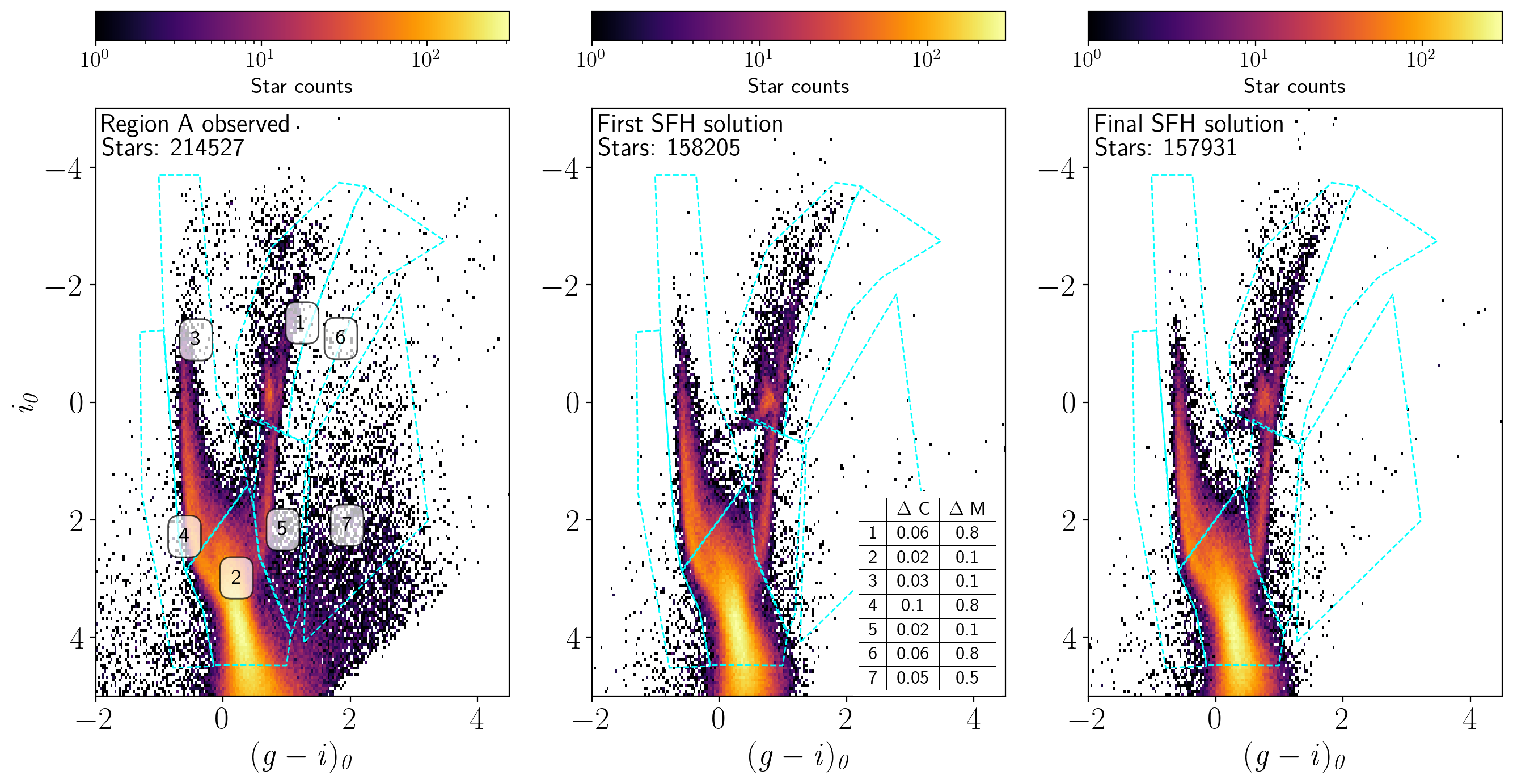}
\includegraphics[scale=0.52]{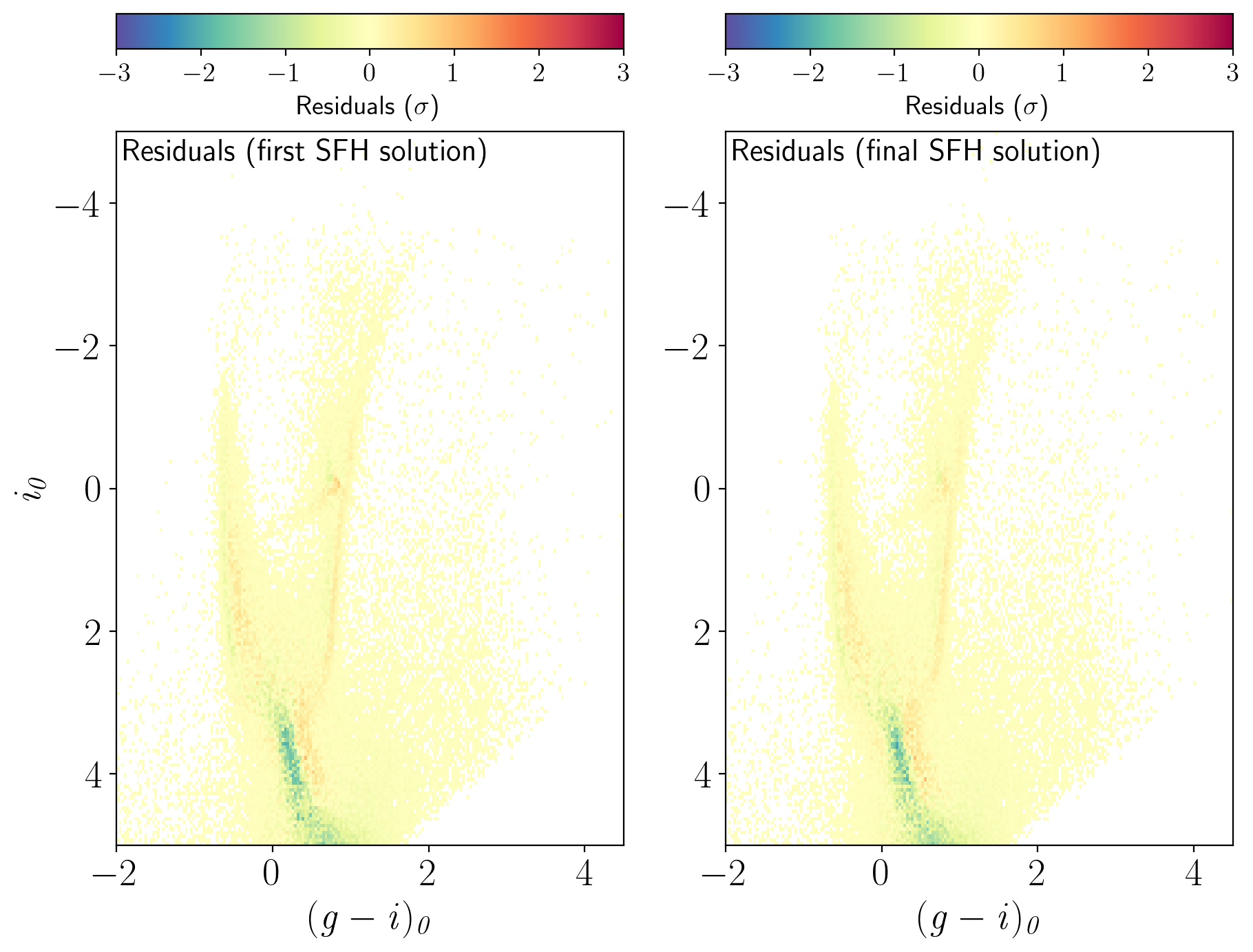}
\caption{Top left: The observed CMD parametrised using the bundle strategy; these bundles were further binned into boxes of varying dimensions which add different weights in the final fit (see text for details). Top middle: The parametrised solution CMD from the \textit{first} SFH derivation with a table showing the binning in colour ($\Delta$ C) and magnitude ($\Delta$ M) applied during the fitting process for all of our results. Top right: Same as top middle, but for the \textit{final} SFH derivation \textit{after} the line-of-sight depth is accounted for. Bottom left: Residual CMD (observed - SFH solution) in units of Poisson sigmas for the \textit{first} SFH solution. Bottom right: Same as bottom left, but for the \textit{final} SFH derivation \textit{after} the line-of-sight depth has been accounted for in the SFH derivation.} 
\label{fig:FigCMDsResiduals}
\end{center}
\end{figure*}

We adopt the best-fitting combination as the SFH of our population and, from this, we derive a `solution CMD' that possesses similar characteristics to the observed CMD. For the comparison of the distribution of stars in the observed CMD and each combination of SSPs, we followed an \textit{a la carte} approach and parametrised the observed CMD into six different sections that we call `bundles' according to the nomenclature used in previous works (e.g., \citealt{Monelli2010a}, \citealt{Ruizlara2018}, \citealt{ruizlara2020}, \citealt{Rusakov2021}, \citealt{Massana2022}). 
The bundles are further divided into smaller boxes (containing around 0-100 stars/box) and only the stars within the bundles' limits are considered for the SFH calculation. We show this approach in the top left panel of Figure \ref{fig:FigCMDsResiduals} that displays the observed CMD of \hl{region A} (see Figure \ref{fig:FigCMDsResiduals} caption for more details) and the solution CMD with the bundles overlapped (top middle). The sizes of the boxes in which each bundle is divided are shown as an inset table in the top middle panel. Additionally, the bundles only include stars which are brighter than the magnitude threshold corresponding to a 50$\%$ completeness level ($i_{0} \sim$ 4.9 mag). To account for the MW foreground contamination, we added an extra bundle (bundle 7) populated exclusively by MW halo stars. Bundle 7 was modelled by \verb|THESTORM| using a field located far from the SMC's main body and thus, dominated by MW stars (SMASH Field 139 from \citealt{nidever2021}). Further tests done to validate our bundle strategy can be found in appendix A of \cite{ruizlara2021}.

\hl{Intrinsic sources of uncertainty in our SFH determination include (i) the effect of binning in colour-magnitude and age-metallicity planes and (ii) the statistical sampling of the observed CMD. We deal with these by shifting our colour-magnitude and age-metallicity grids as well as resampling the observed CMDs (as extensively described in \citealt{Hidalgo2011, Rusakov2021}).} In the top middle and bottom left panels of Figure \ref{fig:FigCMDsResiduals} we present the resulting solution CMD and the residual CMD, respectively, for the first SFH derivation. While small residuals are seen across the entire CMD, the most conspicuous residuals are across the MS and the RC. The strongest residuals arise from the use of data that is less than 90\% complete (i.e. stars fainter than $i_{0}\sim$ 3.1 mag) and due to the line-of-sight depth effects. The line-of-sight depth effects are most conspicuously seen in the RC region of the CMD given the RC feature occupies a narrow magnitude range. In Section \ref{subsec:sfhsolvelos} we will discuss this feature further and how we account for this effect in the SFH determination. 

\subsection{Incorporation of line-of-sight depth effects in the SFH determination (step 2)}
\label{subsec:sfhsolvelos}

\begin{figure}
\begin{center}
\includegraphics[scale=0.215]{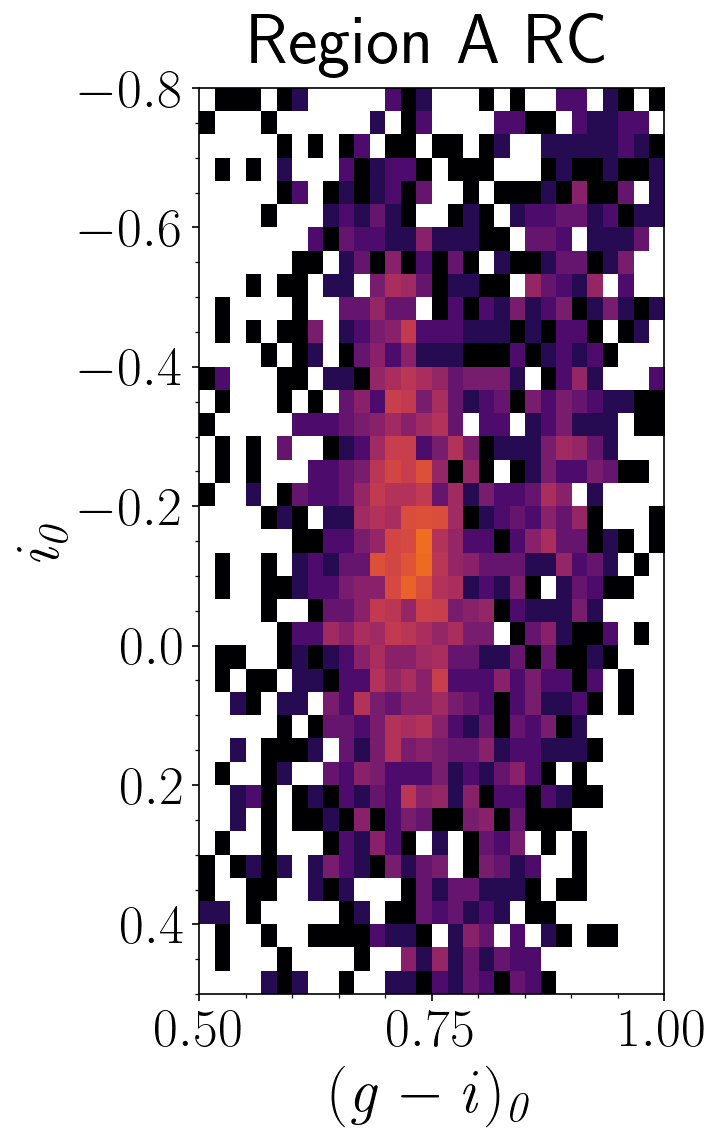}
\includegraphics[scale=0.215]
{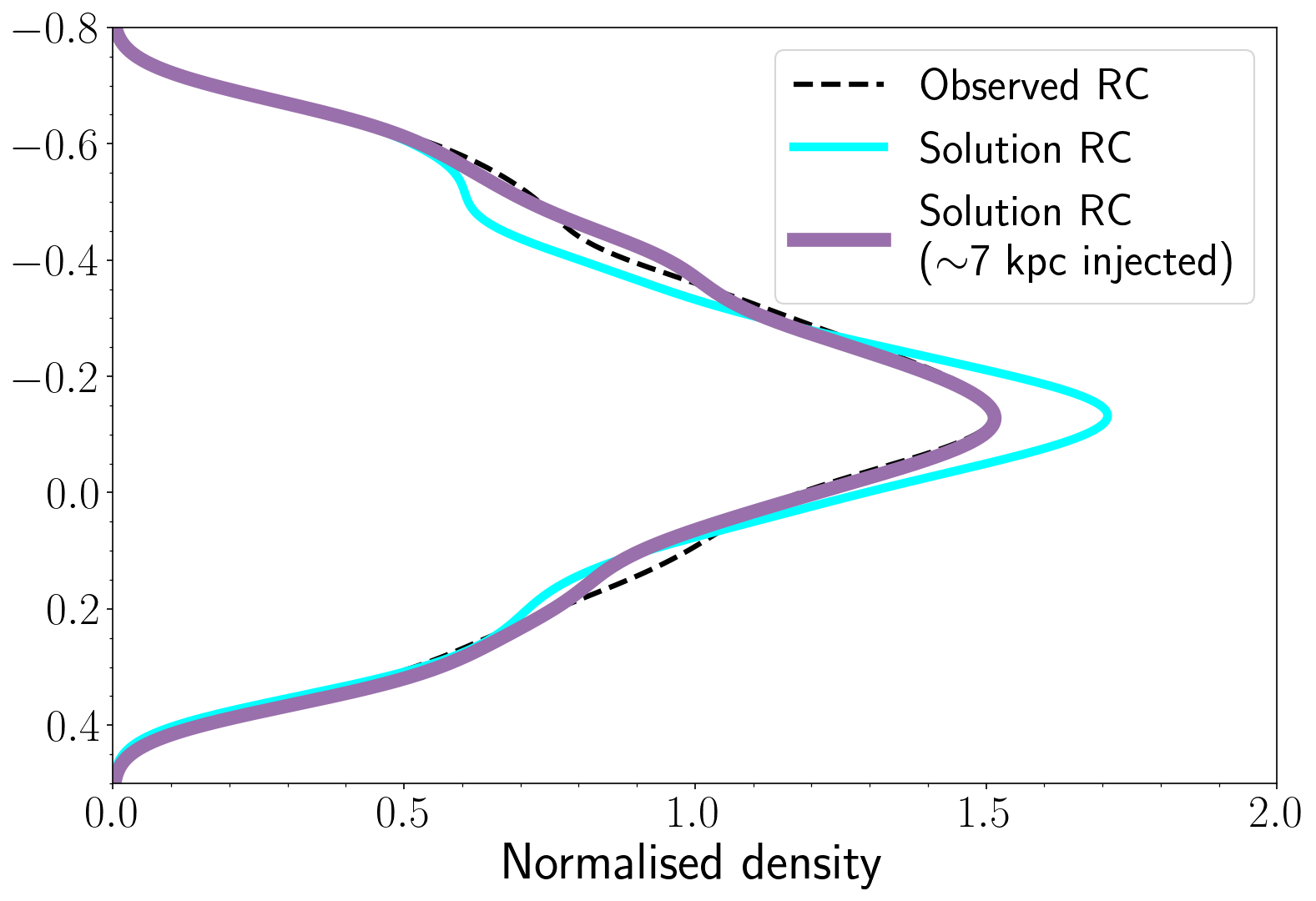}

\caption{Left: zoom in on the red clump region of the observed CMD for \hl{Region A}. Right: luminosity function (represented as a normalised density distribution) of \hl{region A}'s solution CMD RC (cyan) versus observed RC luminosity (black, dashed). In purple is the luminosity function of the solution CMD RC (cyan), \textit{with a line-of-sight depth injected into the CMD} during the line-of-sight depth estimation process. Here, we show the RC luminosity function (purple) with the best-fitting line-of-sight depth found for the region. See text for details.}
\label{fig:RChists}
\end{center}
\end{figure}

The standard procedure to derive the SFH described in Section \ref{subsec:sfhsolve} assumes that all of the stars in the observed sample are at the same distance. This assumption works well for galaxies where the distance among their stars is small in comparison to the distance to the galaxy. This, however, is not the case for the SMC given its line-of-sight depths measure up to approximately $\sim$1/3 ($\sim$20 kpc in the northeast) of its distance to us ($\sim$61 kpc). Therefore, the spread in distance among the SMC's stars is visible in the form of magnitude variations on the RC structure that is expected to be a narrowly concentrated (in magnitude) feature on the CMD\footnote{The RC also shows some intrinsic spread in magnitude. This intrinsic spread depends on population effects- i.e., on the characteristic age and metallicity spread of the underlying stellar population- as well as on the adopted photometric passbands (the spread significantly decreases when moving from the optical to the near-infrared bands. See \citealt{Cassisi2013}, and references therein, for a detailed discussion on this topic).}. 

In order to account for the line-of-sight depth effect in the CMD fitting, we follow an approach in which we apply a magnitude dispersion across the synthetic CMD to simulate the spread in distance. In this Section, we will describe how we quantify the line-of-sight depth comparing the luminosity function of the observed RC with that of the best fitting model (see Sect.~\ref{subsec:sfhsolve}) of the first run of \verb|THESTORM| on the observed CMDs not considering any line-of-sight depth (standard procedure). Then, we simulate the derived line-of-sight depth in the synthetic CMDs that we use to derive the second and final SFH.

\subsubsection{Estimating the line-of-sight depth from the RC luminosity function}
\label{subsubsec:sfhsolveRC}
\hl{Figure \ref{fig:RChists}, left, shows a zoom-in of the RC from the region A's CMD depicted in Figure \ref{fig:FigCMDs}. From the CMD we calculate the observed RC luminosity function (black, dashed histograms, right of Figure \ref{fig:RChists}) and compare it with the `solution' RC luminosity function (cyan histogram in Figure \ref{fig:RChists}). The solution RC luminosity function is derived from the first SFH solution, which gives us the first estimate of the age and metallicity distribution of the stars in the observed RC.} The mismatch between the observed and the solution RC luminosity functions is mostly due to the line-of-sight depth affecting the observed data and not the model (as \verb|THESTORM| was run without simulating any line-of-sight depth to the synthetic CMD). Then, by simulating several degrees of distance spread to the solution CMDs and comparing the resulting RC luminosity function to the observed one, allows us to quantify the distance spread of the stars observed \hl{in the studied region}. First, we assume that the stars in the SMC are located along the line-of-sight following a Gaussian distribution. Assuming a distance modulus of 61.5 kpc as the mean, we construct the Gaussian distribution of stellar distances with its full-width at half maximum (FWHM) representing the total extent of the line-of-sight depth \citep{nidever2013}. After this, we randomly sample the fractional distance modulus shifts from the distribution (ranging in FWHMs from 0 to 23.5 kpc, in steps of 0.2 kpc, 250 times) and inject such shifts, in the form of magnitude shifts, into the absolute magnitudes of solution CMD stars. At each step, we re-measure the solution RC luminosity functions. In order to identify the RC luminosity function with a distance spread that best replicates the observed RC luminosity function, we use a two-sample Kolmogorov-Smirnov (KS) test (\citealt{Chakravarti}) and select the depth with the highest probability statistic. 

\hl{In Figure \ref{fig:RChists}, as a purple histogram, we show the above mentioned best-fitting RC luminosity function measured for region A during our line-of-sight depth estimation procedure. The luminosity function is measured from the solution CMD (from the first SFH run, cyan histogram) after simulating a line-of-sight depth corresponding to the best-fitting estimation. There is now a better agreement between the observed and RC luminosity function. This procedure was carried out for region A and region B, measuring a line-of-sight depth effect of $\sim$7 kpc present in both CMDs.}

\subsubsection{Line-of-sight depth as given by \hl{red clump stars}, variable stars \hl{and star clusters in the northeastern shell}}
\label{subsubsec:sfhsolvevariablestars}

As a sanity check, we cross-matched the SMASH data for region A with 76 CCs and RR-Lyrae stars from the OGLE-IV survey (\citealt{Jacyszyn-Dobrzeniecka2016}, \citealt{Jacyszyn-Dobrzeniecka2017}). We found that, \hl{for region A}, the distribution of CCs and RR Lyrae stars can indeed be represented by respective Gaussian distributions, measuring the line-of-sight depth (the FWHM of the distribution) between $\sim$4.7 kpc (CCs) and $\sim$7.6 kpc (RR Lyrae). Combining the two distance indicators produces a distribution with a line-of-sight depth measurement of $\sim$5.7 kpc.  This value is compatible with our estimate for the line-of-sight depth of region A from the RC luminosity function ($\sim$7 kpc).

\hl{\cite{Piatti2022} (introduced earlier) also derived distances to star clusters on the northeastern shell and surrounding regions, finding a separation of $\sim$15 kpc between the 3 star clusters (NGC 458, HW 64, IC655) that lie within region A. Indeed, if we quote our line-of-sight depth measurement for region A as the range of the best-fitting Gaussian distribution used (approximately 6$\sigma$, which statistically considers 98$\%$ of the stars) rather than its FWHM, then the new depth is 18 kpc, which is compatible with the estimate using star clusters. Given that the northeastern shell occupies a relatively small area of the SMC we are limited by the dearth of variable stars and star clusters. The advantage of using the RC luminosity functions to estimate distances is that the stellar populations of the region at hand are better represented and that the RCs are very well populated in the CMD (\citealt{SubramanianSubramaniam2012}; \citealt{nidever2013}; \citealt{Tatton2021}; \citealt{ElYoussoufi2021}). Our result thoroughly agrees with \cite{SubramanianSubramaniam2012}, who measure line-of-sight depths of 6–8 kpc in the northeastern SMC. We also find good agreement with \cite{Tatton2021}, who quote a line-of-sight depth of either $\sim$17 - 20 kpc or $\sim$5.9 - 7.9 kpc (depending on if the range or 50$\%$ of the near-IR luminosity function is used for their line-of-sight depth measurement) for a northeastern region which includes the northeastern shell and surrounding areas. While $\sim$17 - 20 kpc agrees with the range of our result ($\sim$ 18 kpc), $\sim$5.9 - 7.9 kpc is $\sim$2 - 4 kpc more than 50$\%$ of our result ($\sim$4 kpc). This small deviation is due to the differences in areal coverage, population effect assumptions and measurement methodologies applied.}
\\
In any case, it is important to highlight that the scope of this paper is not a geometrical characterisation of the SMC, but rather to improve the determination of SFHs of the SMC by taking into account the possible effects of the spread in stellar magnitudes caused by the significant distance spread in the SFH solutions.

\subsubsection{Simulating the line-of-sight depth in synthetic CMDs}\label{subsubsec:sfhsolve2ndsfh}

We are now armed with the tools to compute the SFH considering the line-of-sight depth. To achieve this, we construct a Gaussian distribution in which its FWHM represents the best-fitting line-of-sight depth. From the Gaussian distribution, we sample distance modulus shifts, converting them to fractional $g, i$ magnitude shifts. We then apply the magnitude shifts into the stars in the synthetic CMDs and proceed to obtain the SFH as described in Section \ref{subsec:sfhsolve}, with the difference that the synthetic CMD  has now the line-of-sight depth simulated. In Figure \ref{fig:FigCMDsResiduals}, top right, we show \hl{region A}'s solution CMD of the final SFH after the line-of-sight depth has been considered and, in the bottom right panel, we show the residual CMD. Here we see that, in comparison to the first SFH solution's residuals (bottom left panel), when accounting for the line-of-sight depth in \hl{region A} the residuals observed across the RC improve. 
\\
While the robustness of \verb|THESTORM| in deriving SFHs has been tested multiple times (e.g., \citealt{ruizlara2020}; \citealt{Rusakov2021}; \citealt{Massana2022}) the effects of line-of-sight depths on SFH recovery using \verb|THESTORM| have not been previously studied. In the Appendix, we assess the SFH recovery with the inclusion of the line-of-sight depths overviewing previous line-of-sight depth tests on SFH recovery from the literature, we show the full (two-step) SFH procedure on mock data, we assess the effects of the line-of-sight depths on mock data recovery in single bursts and complex SFHs, and show the age-metallicity relations (AMR) for the various scenarios. From all these tests we can conclude that there is evidence that the SFH recovery improves if we consider the line-of-sight depth effects while solving for the SFH. In what follows, and unless otherwise stated, the shown SFHs have been computed using this two-step SFH recovery, i.e., considering the line-of-sight depth effects.

\section{Star formation history results}
\label{sec:sfh}

\begin{figure*}
\begin{center}
\includegraphics[scale=0.5]{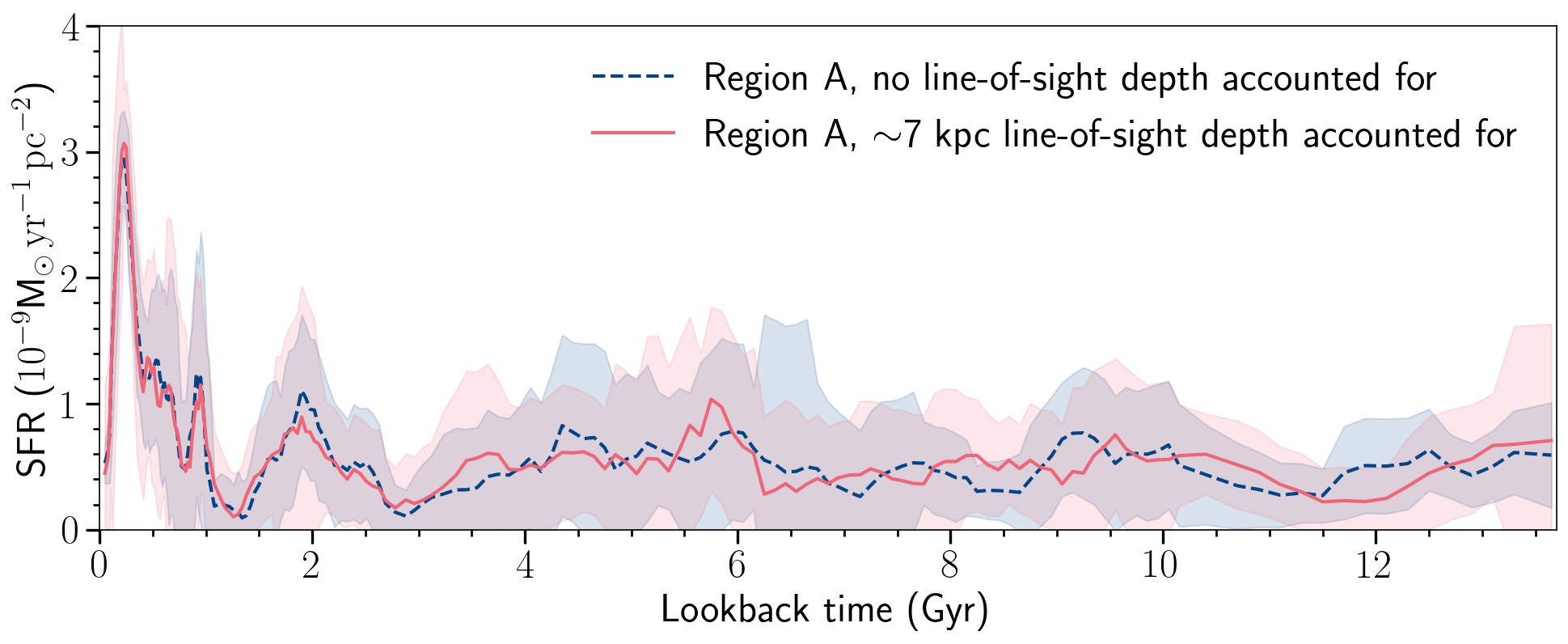}
\includegraphics[scale=0.5]{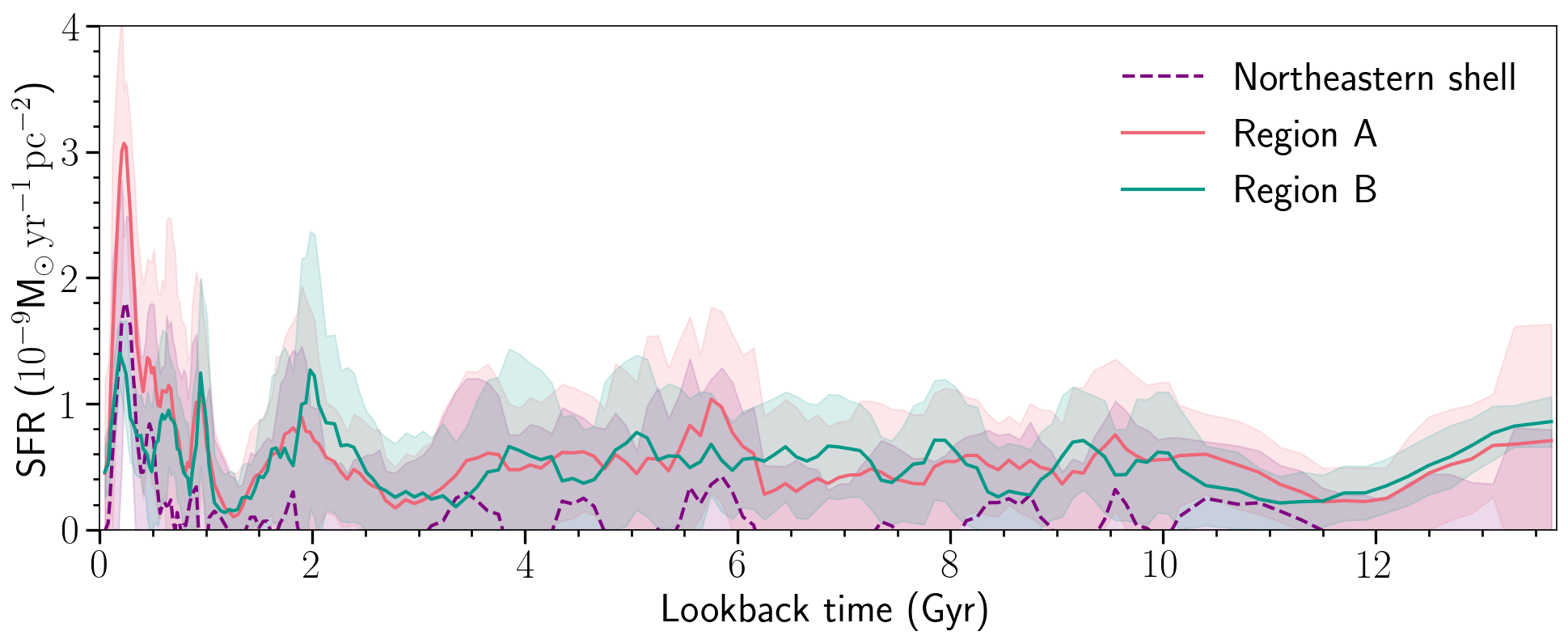}
\caption{Top: SFR(t) results for \hl{region A} before and after we account for the line-of-sight depth during the SFH derivation process. Bottom: SFR(t) results for \hl{region A, region B, and the final northeastern shell SFR(t) (after the subtraction of the SFR(t) contribution of region B to the SFR(t) of region A)}. The results shown are obtained after accounting for the line-of-sight depth during the SFH derivation process. Shaded areas represent uncertainties as explained in the text.}\label{fig:FigSFHs}
\end{center}
\end{figure*}

\begin{figure*}
\begin{center}
\includegraphics[scale=0.5]{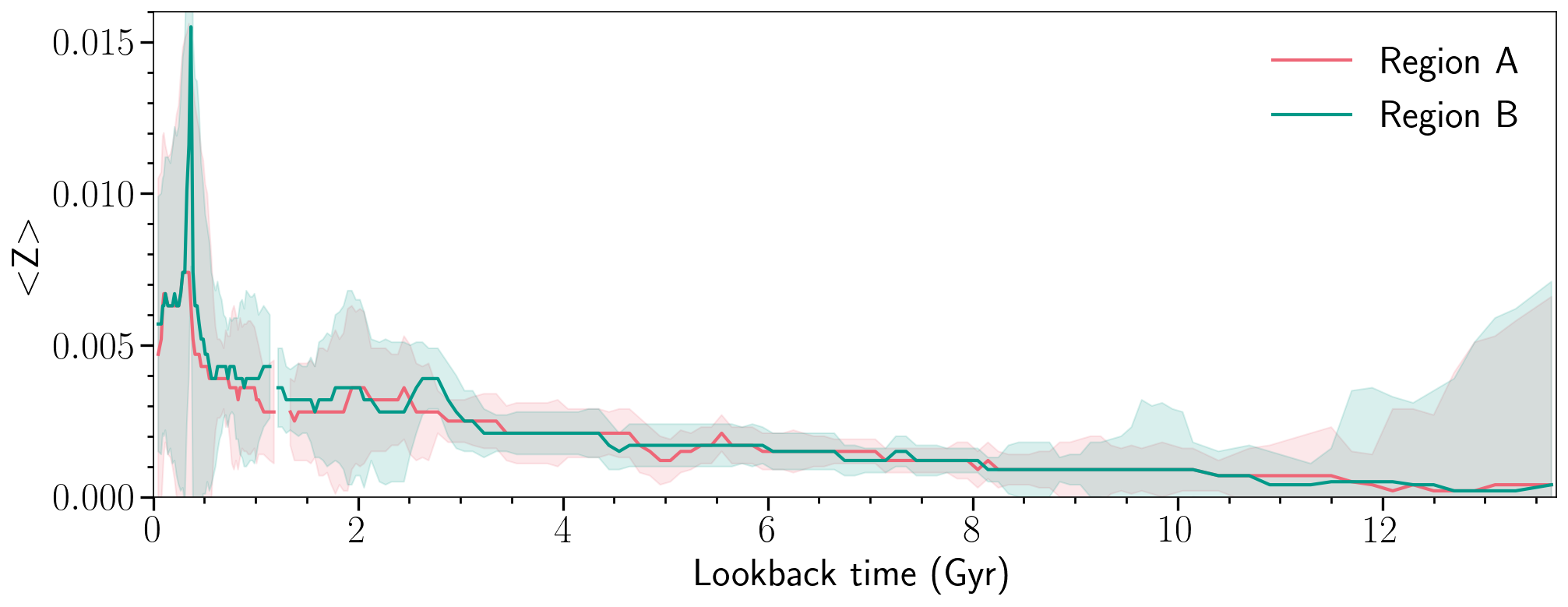}
\caption{Age-metallicity relations displaying the mean metallicity (depicted as <Z>) over lookback time for \hl{region A and region B} after accounting for the line-of-sight depth during the SFH derivation process. Shaded areas also represent uncertainties. Here we mask the age bins in which SFR is less than 5$\%$ of the maximum SFR for the region to avoid uncertain metallicity determinations driven by the lack of stars of such ages.}\label{fig:FigZs}
\end{center}
\end{figure*}

\hl{The SFHs, understood as the star formation rate (SFR) and chemical enrichment (mean metallicity, depicted as <Z>) as a function of time, recovered for the studied regions are presented in Figure \ref{fig:FigSFHs} and Figure \ref{fig:FigZs}. In top of Figure \ref{fig:FigSFHs} we show the SFR(t) for region A before and after the simulation of the line-of-sight depth. Both SFHs are in good agreement within the errors showing recent young ($\le \sim$1 Gyr) star formation with enhancements at $\sim$1 Gyr, $\sim$0.65 Gyr, $\sim$0.45 Gyr, and a recent, conspicuous peak at $\sim$0.25 Gyr, which is $\sim$2-3 times as intense as the other young enhancements. Indeed, the agreement between the SFH results before and after the line-of-sight depth has been considered reinforces the robustness of our method in deriving the SFHs of the SMC. At intermediate ages, there is a peak in the SF at  $\sim$2 Gyr ago in agreement with the findings from \cite{Massana2022} for the whole SMC body. The rest of the SFH does not present clear enhancements in the SF at intermediate-age and old ages. One explanation for this could be the decreased sensitivity for resolving short bursts at older ages as presented in the Appendix, where we show that, beyond $\sim$3.5 Gyr, our ability to resolve short ($\sim$0.1 Gyr), not prominent bursts of star formation decreases due to the photometric limitations of our dataset and the progressively low intrinsic age resolution towards the older ages. In the rest of the Figures, the SFRs and mean metallicities had the line-of-sight depth simulated.}

\hl{The bottom panel of Figure ~\ref{fig:FigSFHs} shows a comparison between the SFH of region A (pink) and the SFH of the control region B (shown in green). The line-of-sight depth measured in the CMD of the control region B was also found to be $\sim$7 kpc. The SFHs follow each other well and are within their respective errors until $\sim$0.5 Gyr ago when the SFH of region A increases outside of the error bars of the SFH from region B. The differences that follow are clearly seen once we subtract the two SFHs and plot the resultant SFH, shown in purple, dashed lines. Now we have recovered the SFH of the northeastern shell (excluding contamination from SMC field stars) and can appreciate the young star formation exclusively belonging to it. The northeastern shell's enhancements at $\sim450$ Myr, $\sim250$ Myr are the most conspicuous in comparison to the rest of its SFH, with the $\sim250$ Myr peak measuring approximately twice the intensity of the $\sim450$ Myr enhancement. All star formation at ages older than $\sim$0.5 Gyr is compatible with zero within the respective errors.}

\hl{In Figure \ref{fig:FigZs} we present the mean metallicity as a function of stellar age for regions A and B. Given that the mean metallicity is calculated as the average of the stars present within the respective age bin, we expect the mean metallicity to fluctuate. In the case of a lack of a population of a given age, the mean metallicity will strongly fluctuate and be uncertain and therefore we mask those areas (regions where the SFR is less than 5$\%$ of the maximum SFR); at younger ages, we use narrower metallicity bins (see Section ~\ref{subsec:sfhsolve}) and therefore we are more likely to see fluctuations at these ages. Both regions A and B show a mostly linear increase in metallicity, with a sharp increase at $\sim$ 0.45 Gyr that coincides with the enhancement of SF at that age (see Figure 6). In addition, the <Z> from $\sim$0.5 Gyr ago to 1.5 Gyr ago for region B is higher than that of region A but given that all of the mean metallicities are still within the error bars, we cannot draw major conclusions from this. We can, however, suggest that both regions A and B follow similar chemical enrichment histories as the northeastern shell (the differentiating factor between the two regions, affecting marginally the overall metallicity).}

\hl{Next, we constructed cumulative mass fractions (CMF) as a function of lookback time. CMFs (also known as cumulative SFHs) indicate which fraction of the present-day stellar mass has formed before a specific lookback time. CMFs are useful as they help circumvent uncertainties in the measured SFHs since we look at total stellar mass assembled up to a given time rather than an instantaneous SFR.}

\begin{figure*}
\begin{center}
\includegraphics[scale=0.32]{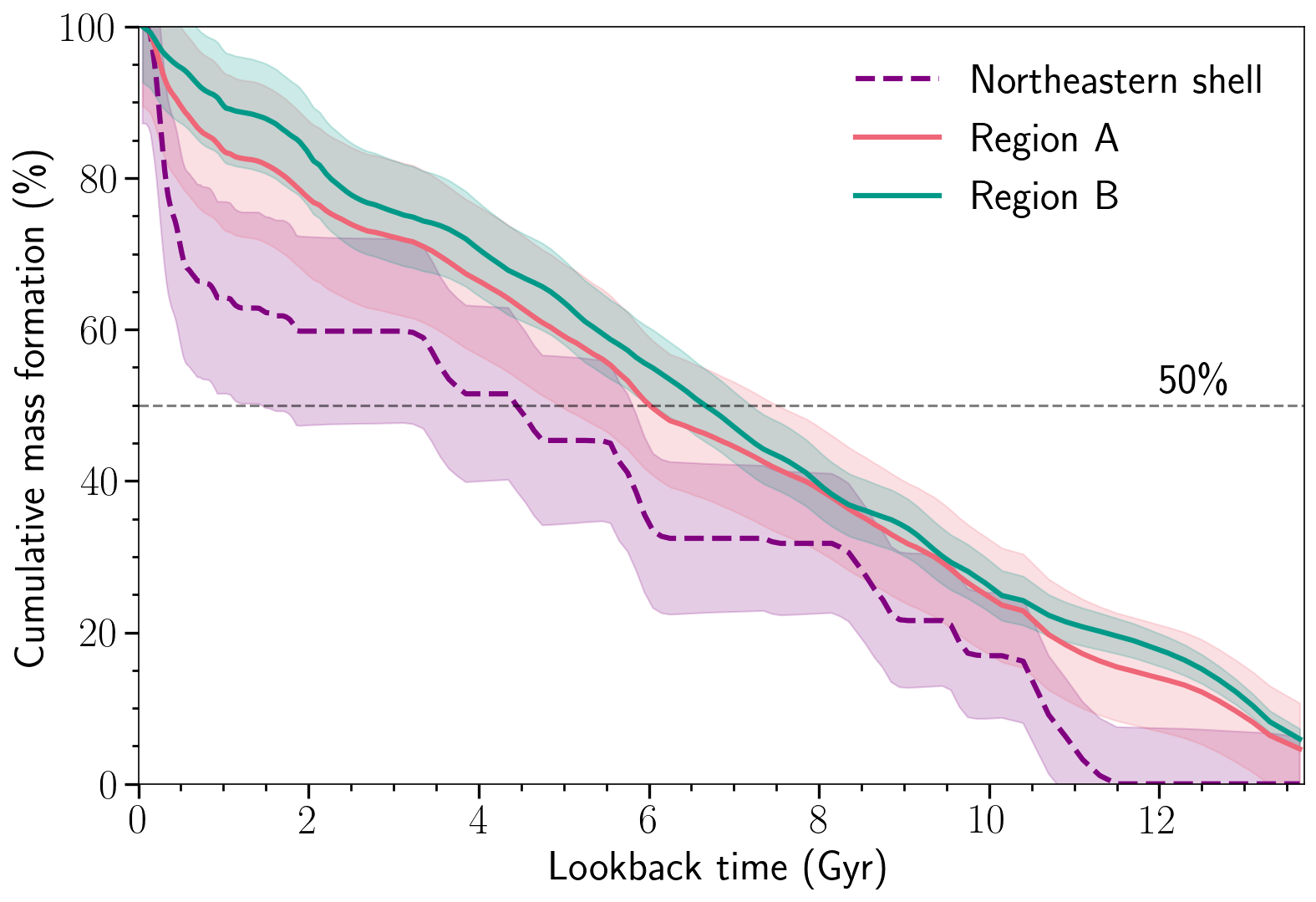}
\includegraphics[scale=0.32]{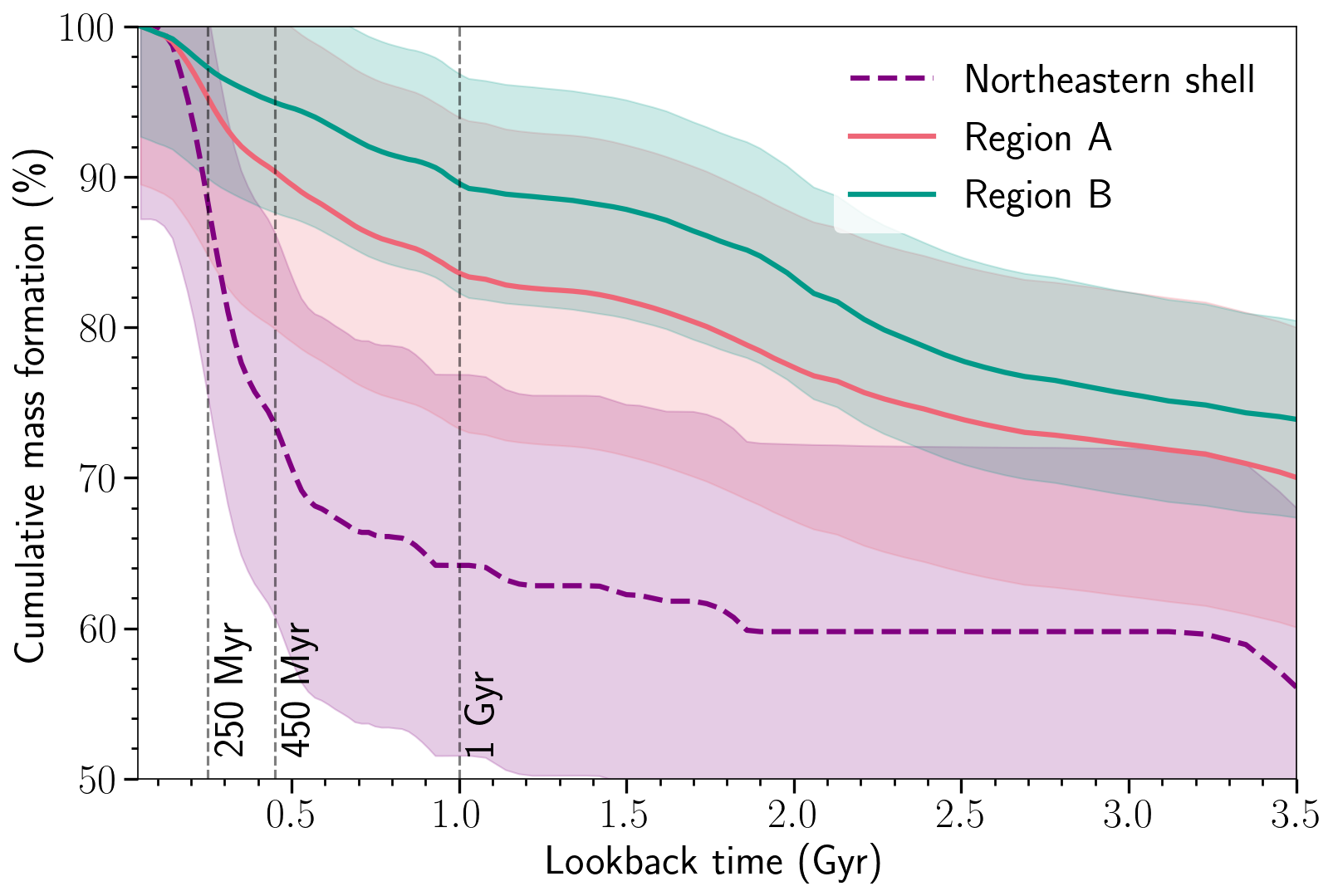}
\caption{Cumulative mass formation (CMF) for \hl{the northeastern shell, region A and region B.} In the left panel, we show the CMFs throughout the whole history of the galaxy; we mark the 50$\%$ CMF. In the right panel, the CMFs from 3.5 Gyr ago up until now are shown; the vertical dashed lines mark the northeastern shell's SF peaks and the start of young star formation (1 Gyr).}\label{fig:FigCMFs}
\end{center}
\end{figure*}

\hl{In the left panel of Figure \ref{fig:FigCMFs} we explore the CMF in region A, region B and the northeastern shell up to 13.7 Gyrs ago, and in the right panel we present the CMF result up to 3.5 Gyrs ago. In both cases we show the CMF of the SFHs after including the line-of-sight depths. The right panel of Figure \ref{fig:FigCMFs} also includes the northeastern shell's star formation bursts and the start of young star formation (1 Gyr) depicted as grey dashed vertical lines with the ages annotated.} 

\hl{The CMF of the northeastern shell indicates that the steepest mass formation gradient occurred within the last $\sim$0.5 Gyr, with at least $\sim$32\% of the total stars formed within the last $\sim$0.5 Gyr, and $\sim$35\% (only 3$\%$ more) within the last $\sim$1 Gyr. This corresponds to a stellar mass of $1.15\pm^{1.15}_{0.86} \times 10^{5}$ M$_{\odot}$ within the last $\sim$0.5 Gyr, and $1.38\pm^{2.46}_{1.09} \times 10^{5}$ M$_{\odot}$ within the last $\sim$1 Gyr. We should highlight that this is a lower limit consequence of our way of computing the cumulative SFH, as age intervals with positive SFR contribute to the total mass of the system despite being consistent with a zero SFR. This exponential rate of mass formation began just under $\sim$1 Gyr ago. This is in stark contrast to region B where the mass formation followed a linear gradient from $\sim$1 Gyr onwards, with only $\sim$10$\%$ of the total mass formed within the last $\sim$1 Gyr. The CMF of region A (which contains the northeastern shell) also indicates an exponential rate of young mass formation, with $\sim$16\% of its stars formed within the last 1 Gyr (corresponding to $3.46\pm^{2.56}_{2.22} \times 10^{5}$ M$_{\odot}$). Finally, the CMFs also show that $\sim$50\% of the stars formed in the northeastern shell over the last $\sim$4 Gyr, whereas 50$\%$ of the mass around the northeastern shell in region B was built nearly $\sim$4 Gyrs earlier.}

Our findings from quantitative SFH determinations are in good agreement with the qualitative age determinations by MD19, who combined information from overlapping isochrones on CMDs and colour functions, and used the age distributions of CCs and clusters within the \hl{northeastern shell. They also agree well with the quantitative star cluster age determinations by \cite{Piatti2022}, who dated the star clusters (corresponding to our areal definition of the northeastern shell) to be between $\sim$44 - 135 Myrs old. While we do not observe conspicuous peaks at these ages, our SFR is active and within the error bars in this age range. Finally, the results presented agree remarkably well with \cite{Hota2024}, who qualitatively overlapped isochrones on CMDs (FUV-optical and optical, respectively) and identified enhancements 40 Myrs and 260 Myrs ago. From our analysis we can confirm that the shell is a young structure, formed mostly (or exclusively) by stars younger than $\sim$500 Myr, with peaks of star formation $\sim$450 and $\sim$250 Myr ago.}

\section{Discussion}
\label{sec:discussion}

\begin{figure*}
\begin{center}
\includegraphics[scale=0.52]{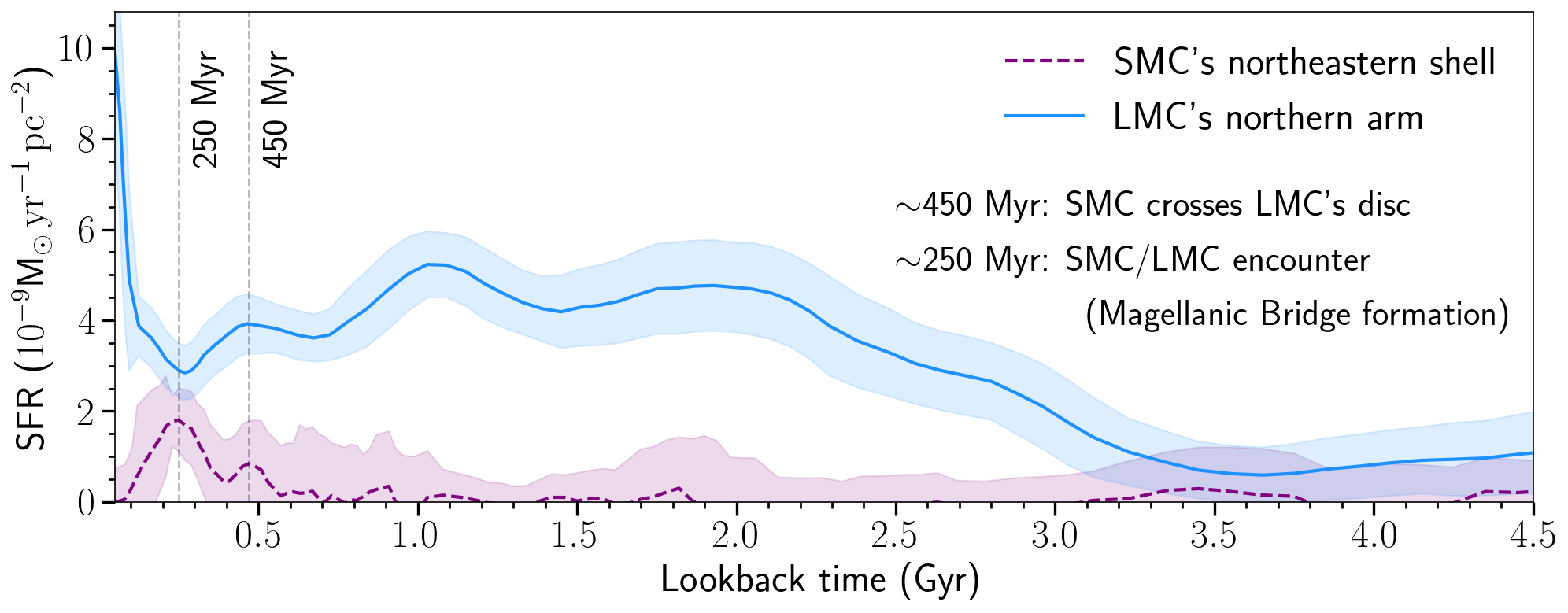}
\caption{Comparison between the \hl{northeastern shell's SFH after region A subtraction} (purple dashed line) and the SFH of the LMC's northern spiral arm (blue line) (see \citealt{ruizlara2020}). The grey vertical dashed lines highlight the shared SFR enhancements in the \hl{northeastern shell }and the LMC's northern arm at $\sim$0.25 Gyr and $\sim$ 0.45 Myr. \hl{These enhancements align with the timing of the SMC crossing the LMC's disc (e.g. \citealt{Cullinane2022, Navarrete2023}) and eventually colliding with the LMC, forming the Magellanic Bridge (e.g. \citealt{Choi2022}).} }\label{fig:FigSFHsLMC}
\end{center}
\end{figure*}

\hl{While the SFRs, mean metallicities, and CMFs presented in Section \ref{sec:sfh} for regions A, B, and the recovered northeastern shell show many similarities (see Figure \ref{fig:FigSFHs}, Figure \ref{fig:FigZs}, and Figure \ref{fig:FigCMFs}), there are some features of interest we highlight in this discussion. These features coincide with possible pericentric passages between the MCs and the increased influence of the MW on the MCs over the last $\sim$500 Myrs (e.g. \citealt{Besla2007, Patel2020}).}

\hl{Beginning $\sim$3 Gyr ago, region A and region B (selected to represent the SMC’s field stars) share a similar SFR pattern, matching in intensity at $\sim$1 Gyr. Following this period, region A’s overall SFR rapidly increases, while region B’s SFR remains relatively constant. From $\sim$500 Myrs onwards, the SFR of region A rises above the error bars of region B. By subtracting the SFH of region B from region A, we derive the northeastern shell’s SFH, revealing significant peaks $\sim$450 Myrs and $\sim$250 Myrs ago. These trends are also evident in the CMFs, with the northeastern shell’s CMF rapidly rising from $\sim$500 Myr ago, resulting in at least $\sim$32\% (lower limit) of the total stars being formed ($1.15\pm^{1.15}_{0.86} \times 10^{5}$ M$_{\odot}$). The stark contrast in the SFRs and CMFs between region A and region B beginning $\sim$500 Myr ago suggests that this enhancement was only significant for the northeastern shell, as it is not as pronounced in the surrounding field stars. The $\sim$250 Myr enhancement is the strongest and the only peak not compatible with zero SF (according to the respective errors). As such, our results strongly suggest the northeastern shell formed within the last $\sim$500 Myrs, with a stellar mass totalling $1.15\pm^{1.15}_{0.86} \times 10^{5}$ M$_{\odot}$. Whether it began forming $\sim$500 Myr ago or solely from the $\sim$250 Myr enhancement alone is difficult to discriminate.}
\hl{While our strategy of subtracting region B’s contribution unveiled the young nature of the northeastern shell, traces of intermediate-age and old populations remain in our SFH results. Whether these belong to the northeastern shell or are part of the field stars not fully removed from region A is extremely difficult to assess. However, based on Figures \ref{fig:FigSFHs} and \ref{fig:FigCMFs}, the intermediate-age and old populations are consistent with a zero SFR. Consequently, the stellar mass formed in the northeastern shell over $\sim$1-2 Gyrs ago should probably be considered an artefact due to uncertainties in deriving SFHs. In other words, our strategy in isolating the northeastern shell’s SFH by subtracting region B’s SFH from region A’s SFH is not exempt from errors, and it is likely that all stellar mass formed prior to $\sim$1-2 Gyrs ago is not real. The only clear signal, where respective errors do not suggest zero star formation at any point, is the star formation over the last $\sim$250 Myr. This supports the idea that the SMC's northeastern shell formed most of its mass $\sim$250 Myrs ago, even if star formation began $\sim$500 Myr ago (or $\sim$1-2 Gyrs ago, but not older than that).}

\hl{In Figure \ref{fig:FigSFHsLMC} we present a comparison between the SFHs of northeastern shell and the SFH of the LMC's northern spiral arm  \citep{ruizlara2020}, also obtained using SMASH photometry. We find that the enhancement in the northeastern shell's SFH $\sim$0.45 Gyrs ago coincides with an enhancement in the LMC's SFH. Following this, the northeastern shell's enhancement $\sim$0.25 Gyr ago coincides with a drought of star formation in the LMC, before the northeastern shell’s SFH sharply declines. This implies a period of intense star formation followed by a decrease for the northeastern shell, as well as a delay in the star formation between the northeastern shell and the LMC (the drought in SF is rapidly followed by an increase), probably consequence of the last interaction between the clouds, $\sim$250 Myr ago \citep[][]{Zivick2018}. This pattern highlights the dynamic nature of the northeastern shell’s formation and evolution within the last $\sim$500 Myrs, with the most substantial star formation occurring $\sim$250 Myrs ago.}

\subsection{Potential scenarios for the formation of the northeastern shell}
\label{subsec:shellorigin}
In recent years, there has been growing evidence that the halo regions of nearby galaxies (up to their virial radii) contain multiphase gas, known as ‘circumgalactic medium’ (CGM; e.g., \citealt{Tumlinson2017}; \citealt{Armillotta2017}).  
Given that it traces the inward flows from the intergalactic medium as well as the outward flows of enriched material from galaxies, the CGM has become an important player in our understanding of how galaxies evolve in their environments. As such, the CGM is a key fuel for setting star formation in a galaxy (e.g., \citealt{Binney1977}; \citealt{Gatto2013}), a venue for galactic feedback and recycling (e.g., \citealt{Hobbs2013}; \citealt{Agertz2020}), and a main regulator of the supply of gas within the galaxy (e.g.,
\citealt{Dave2012}; \citealt{Fraternali2013}).
The proximity to the MW is believed to be the reason for the removal of material from nearby galaxies (e.g., \citealt{Grcevich2009}; \citealt{Gatto2013}), so the interactions between the MC system and the MW's CGM likely alter the appearance and dynamics of their gaseous components. As the MCs fall through our Galaxy's CGM, tidal forces strip loosely bound gas from their discs, which then would form part of the massive Magellanic Stream \citep{DOnghia2016}. Indeed, simulations that include CGM surrounding the LMC and SMC can explain the ionised gas component of the Magellanic Stream \citep{Lucchini2020}.
\\
Tidal interactions unquestionably play a dominant role in shaping the gas in the MC system such as in the case of the Magellanic Bridge region where observations (\citealt{Noel2013}; \citealt{Noel2015}; \citealt{Carrera2017}; \citealt{Grady2021}) attribute the significant amount of gas and stars found there to tidal forces from encounters between the MCs, a fact also confirmed by current models (e.g., \citealt{Besla2010}; \citealt{Besla2012}). On the other hand, ram pressure stripping has in all likelihood contributed to the structure of the MCs gaseous discs (e.g., \citealt{Mastropietro2005, Salem2015, Lucchini2021}). This is mainly supported by the fact that, while the bulk of the Magellanic Stream has a very low metallicity, the parts closer to the MCs are more metal-rich \citep{Richter2013} suggesting that it has possibly been enriched from gas that has been stripped from the MCs \citep{Lucchini2020} via ram pressure. Moreover, the LMC possesses a northeastern edge, also called the `leading-edge' \citep{Salem2015}, that shows an abrupt truncation in the HI gas at $\sim$6.2 kpc from the LMC's centre but with a stellar profile that continues uninterruptedly well beyond this radius \citep{vanderMarel2001}.
\\
Following the above discussion, the MCs have conceivably experienced an increase in gas removal due to the introduction of ram pressure from the MW's CGM. In addition, there is a gas wake behind the LMC and a large bowshock surrounding the LMC-SMC system caused by the passage of the LMC through the MW CGM, where the SMC is predicted to have spent the last $\sim$0.25 Gyr in or near the bowshock (\citealt{Salem2015, Setton2023}). As a result of these factors, the SMC is expected to have recently travelled through a higher density gas environment than it has in the past. Given that ram pressure is known to increase SF on the leading edges of infalling galaxies (\citealt{Verdugo2015}), and assuming that the leading edge of the SMC's orbit is located at the northeastern part of it (see discussion below), this may explain the increased intensity of the SF burst in the SMC's northeastern part of the SMC (giving way to the shell) $\sim$0.25 Gyr ago. Furthermore, ram pressure has been suggested to both trigger and quench star formation in dwarf-dwarf galaxy interactions (\citealt{Kado-Fong2023})- this could have contributed to the simultaneous trough in the LMC's SFR $\sim$0.25 Gyr ago and SF enhancement in the SMC's northeastern shell. After $\sim$0.25 Gyr, as the SMC continued to approach the LMC's disc, the SMC's gas may have been stripped and the LMC's gas compressed and SF reignited.
\\
We suggest that one possibility for the formation of the northeastern shell is that, as the SMC was approaching the LMC's disc and the bowshock over the past $\sim$0.5 Gyr, its gas was shocked into a shell of star-forming gas. Then, the gas shell experienced SF enhancements in line with the SMC's orbit around the LMC, and the intensity of the last SF burst $\sim$250 Myr could be explained by the SMC passing through an increasingly dense environment approaching the bowshock before its close encounter with the LMC. In fact, these bowshocks have been linked to the LMC's large "shell" of dense star-forming gas \citep{deboer1998}.

\hl{On the other hand, our results are also compatible with another scenario in which the tidal stripping of gas and stars from the inner SMC during the formation of the Magellanic Bridge $\sim$250-150 Myr ago could have formed the northeastern shell (e.g., \citealt{Piatti2022}), and contributed to the young, intermediate-age, and old stellar populations detected. This could still be the case even if the ages and metallicities of the field stars around the northeastern shell are similar. However, a purely tidal origin is unfeasible given that, in a tidal interaction scenario, intermediate-age and old populations are also stripped \citep{Noel2013}, and these are not clearly visible to trace the northeastern shell in Figure \ref{fig:FigRegions}. As previously discussed, the intermediate-age and old populations detected in the SFHs derived are compatible with zero star formation. In addition, \cite{Hota2024} find that the proper motions and velocity dispersion of the northeastern shell's $<$400 Myrs old stars do not show signs of significant tidal stripping.}

\hl{We have to be cautious in arriving at conclusions here because, unlike the LMC, the SMC is surrounded by a large amount of HI in all directions, with no evidence of a 'leading-edge' analogue, since there is no location in the SMC where the gas is truncated but not the stellar profile. Although we know that the SMC is crossing over the CGM at high speeds (e.g., \citealt{Costa2011}), its position, orientation, and motion are considerably less well-constrained than for the LMC. For instance, the past pericentric approach of the SMC to the MW is highly uncertain due to its orbit being strongly perturbed by the LMC. Unfortunately, current hydrodynamic simulations do not have sufficient resolution to predict stellar structures such as the northeastern shell, so we cannot unequivocally confirm or rule out the frameworks discussed here.}

\section{Conclusions}
\label{sec:conclusions}
We presented here the quantitative SFH analysis of the SMC's northeastern shell. \hl{To isolate its stellar content, we first selected a region (region A) encompassing the northeastern shell and a control region around it (region B), representative of the field population in the northeastern SMC close to the shell. The northeastern shell's SFH was successfully obtained by subtracting the SFH of region B from region A.} We incorporated, for the first time, the SMC's line-of-sight depth in the SFH calculation. We derived it by comparing the observed RC luminosity function against the theoretical RC luminosity function (from a first SFH obtained in a standard manner, without considering line-of-sight depth) after simulating in it different line-of-sight depths. We then injected the best fitting line-of-sight depth into our model and re-calculated the final SFH. We decomposed the SFH into SFR(t), <Z>(t) and the CMFs for \hl{regions A and B and the northeastern shell}. Finally,  we compared the northeastern shell's SFH with the SFH of the LMC's northern arm (obtained using SMASH) to discuss potential frameworks for the northeastern shell's origin. We summarise our overall findings below:

\begin{itemize}

    \item \hl{The line-of-sight depth measured in the CMDs of both regions was quantified at $\sim$7 kpc. This is in agreement with depth estimates in the northeastern SMC also using RC stars (e.g., \citealt{SubramanianSubramaniam2012}; \citealt{Tatton2021}).} 
    \item \hl{The mock tests presented demonstrate that accounting for the SMC's line-of-sight depth effects results in a more accurate SFH derivation, especially in the metallicity recovery. The improvement becomes more significant with increasing line-of-sight depth and it is especially the case for large line-of-sight depths ($\sim$21-22 kpc). For smaller depths such as $\sim$7 kpc the SFHs recovered before and after simulating the depth maintain agreement with each other and do not show significant changes. Despite this, decreased residuals during the CMD fitting are observed, suggesting that accounting for the line-of-sight depth for region A (and region B) during the SFH procedure is worthwhile.}
    \item \hl{The comparison between the northeastern shell's SFH and the LMC's northern arm SFH shows a synchronous enhancement in both $\sim$450 Myr ago, and a northeastern shell enhancement $\sim$250 Myrs ago coinciding with a trough in the northern arm's SFH $\sim$250 Myrs ago. We therefore link the formation of the northeastern shell to the interaction history of the SMC with the LMC/MW during the past $\sim$500 Myrs}.
    \item \hl{After subtracting the contribution from the SMC's field stars, we find that the northeastern shell is mainly composed by young stars (mostly younger than $\sim$500 Myrs), exhibiting conspicuous, young SF enhancements at $\sim$250 Myr and $\sim$450 Myr. While we also detect small contributions of stars older than $\sim$500 Myrs, this is highly likely to be due to the remaining contamination within the CMD.}
    \item \hl{Finally, we argue that the complex processes of ram pressure stripping and CGM interplay involved in the SMC/LMC/MW system over the past $\sim$500 Myr may have also played an important role in the formation of the northeastern shell. To help assess if the physical conditions from $\sim$500 Myrs ago until now ago could have led to the formation of a shell-like structure similar to the one studied, a higher resolution hydrodynamical model of the SMC/LMC/MW system is needed.}   
    
\end{itemize}

In summary, we have demonstrated our ability to obtain the SFH of the SMC's northeastern shell in unprecedented detail using a novel
technique incorporating line-of-sight depth effects during the SFH computation. Regarding the formation of this structure, \hl{we favour a scenario where the shell formed over the last $\sim$500 Myr. However,} better constraints on the SMC's position, orientation, and motion are needed. In order to draw further conclusions, improved models of the past pericentric approach of the SMC to the MW, and of the SMC's orbit around the LMC, are required. Such models should, ideally, address the high uncertainties in the SMC's orbit due the perturbing effects of the LMC and MW.

\section*{Acknowledgements}

We thank the anonymous referee for their insightful comments which helped improve the quality of the manuscript. JDS acknowledges funding from the Bell Burnell Graduate Scholarship Fund (BB003) and the Engineering \& Physical Sciences Research Council (EPSRC). JDS also thanks the University of Granada for its hospitality, and funding from both the ERASMUS+ mobility scheme and the Surrey-Santander PhD Travel Award.
TRL acknowledges support from Juan de la Cierva fellowship (IJC2020-043742-I) and support from project PID2020-114414GB-100, financed by MCIN/AEI/10.13039/501100011033. CG acknowledges support from the Agencia Estatal de Investigación del Ministerio de Ciencia e Innovación (AEI-MCINN) under grant “At the forefront of Galactic Archaeology: evolution of the luminous and dark matter components of the Milky Way and Local Group dwarf galaxies in the Gaia era” with reference PID2020-118778GB-I00/10.13039/501100011033 and from the Severo Ochoa program through CEX2019-000920-S. SC acknowledges support from PRIN MIUR2022 Progetto "CHRONOS" (PI: S. Cassisi) finanziato dall’Unione europea - Next Generation EU. DMD acknowledges financial support from the Talentia Senior Program
(through the incentive ASE-136) from Secretar\'\i a General de 
Universidades, Investigaci\'{o}n y Tecnolog\'\i a, de la Junta de
Andaluc\'\i a. DMD acknowledges funding from the State Agency for
Research of the Spanish MCIU through the ``Center of Excellence Severo Ochoa" award to the Instituto de Astrof{\'i}sica de Andaluc{\'i}a (SEV-2017-0709) and project (PDI2020-114581GB-C21/ AEI / 10.13039/501100011033). M.M. acknowledges support from the Agencia Estatal de Investigaci\'on del Ministerio de Ciencia e Innovaci\'on (MCIN/AEI) under the grant "RR Lyrae stars, a lighthouse to distant galaxies and early galaxy evolution" and the European Regional Development Fun (ERDF) with reference PID2021-127042OB-I00, and from the Spanish Ministry of Science and Innovation (MICINN) through the Spanish State Research Agency, under
Severo Ochoa Programe 2020-2023 (CEX2019-000920-S).
This paper includes data based on observations at Cerro Tololo Inter-American Observatory, NSF's National Optical-Infrared Astronomy Research Laboratory (NOAO Prop. ID: 2013A-0411 and 2013B-0440; PI: Nidever), which is operated by the Association of Universities for Research in Astronomy (AURA) under a cooperative agreement with the National Science Foundation.
IRAF is distributed by the National Optical Astronomy Observatory, which is operated by the Association of Universities for Research in Astronomy (AURA) under a cooperative agreement with the National Science Foundation.
This project used data obtained with the Dark Energy Camera (DECam), which was constructed by the Dark Energy Survey (DES) collaboration. Funding for the DES Projects has been provided by the U.S. Department of Energy, the U.S. National Science Foundation, the Ministry of Science and Education of Spain, the Science and Technology Facilities Council of the United Kingdom, the Higher Education Funding Council for England, the National Center for Supercomputing Applications at the University of Illinois at Urbana-Champaign, the Kavli Institute of Cosmological Physics at the University of Chicago, Center for Cosmology and Astro-Particle Physics at the Ohio State University, the Mitchell Institute for Fundamental Physics and Astronomy at Texas A\&M University, Financiadora de Estudos e Projetos, Funda\c{c}\~ao Carlos Chagas Filho de Amparo, Financiadora de Estudos e Projetos, Funda\c{c}\~ao Carlos Chagas Filho de Amparo \`a Pesquisa do Estado do Rio de Janeiro, Conselho Nacional de Desenvolvimento Cient\'ifico e Tecnol\'ogico and the Minist\'erio da Ci\^encia, Tecnologia e Inova\c{c}\~ao, the Deutsche Forschungsgemeinschaft and the Collaborating Institutions in the Dark Energy Survey. The Collaborating Institutions are Argonne National Laboratory, the University of California at Santa Cruz, the University of Cambridge, Centro de Investigaciones En\'ergeticas, Medioambientales y Tecnol\'ogicas-Madrid, the University of Chicago, University College London, the DES-Brazil Consortium, the University of Edinburgh, the Eidgen\"ossische Technische Hochschule (ETH) Z\"urich, Fermi National Accelerator Laboratory, the University of Illinois at Urbana-Champaign, the Institut de Ci\`encies de l'Espai (IEEC/CSIC), the Institut de F\'isica d'Altes Energies, Lawrence Berkeley National Laboratory, the Ludwig-Maximilians Universit\"at M\"unchen and the associated Excellence Cluster Universe, the University of Michigan, the National Optical Astronomy Observatory, the University of Nottingham, the Ohio State University, the University of Pennsylvania, the University of Portsmouth, SLAC National Accelerator Laboratory, Stanford University, the University of Sussex, and Texas A\&M University.

\section*{Data Availability}

 The photometry of SMASH used in this study can be accessed at \url{https://datalab.noirlab.edu/smash/smash.php}. This research uses services or data provided by the Astro Data Lab at NSF’s NOIRLab. NOIRLab is operated by the Association of Universities for Research in Astronomy (AURA), Inc. under a cooperative agreement with the National Science Foundation. All data products can be provided upon reasonable request to the corresponding author.



\bibliographystyle{mnras}
\bibliography{mnras_template} 


\appendix
\section{Assessing the recovery of the SFH: effects of accounting for the line-of-sight depth}
\label{sec:sfhmocksappendix}

The ability of \verb|THESTORM| to recover SFHs has been robustly tested in the literature (e.g, \citealt{ruizlara2020}; \citealt{Rusakov2021}; \citealt{Massana2022}).
In this appendix, we add a new layer and examine the effectiveness of our novel approach in recovering the SFH with and without considering the line-of-sight depth according to our two-step procedure described in Section ~\ref{sec:sfhprocedure}. For this, we perform tests with two different mock populations, described below.

\begin{figure*}
\begin{center}
\includegraphics[scale=0.25]{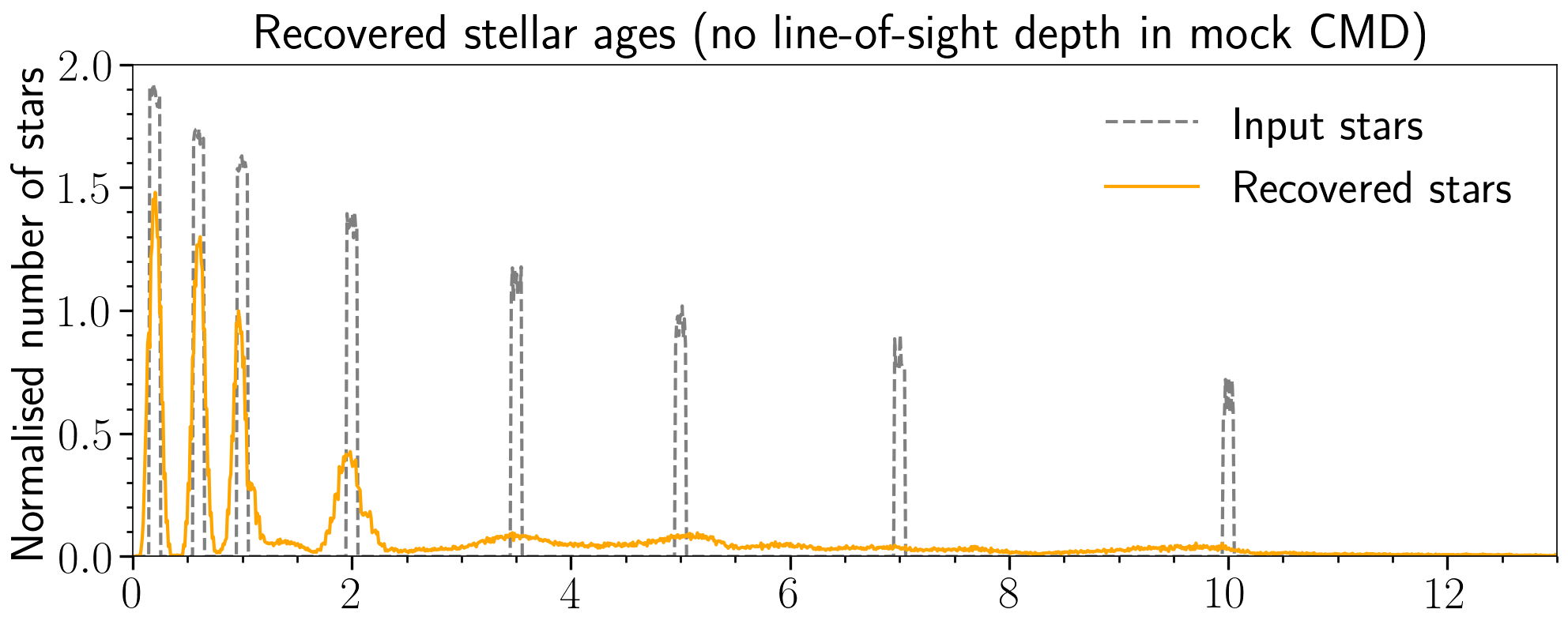}
\includegraphics[scale=0.25]{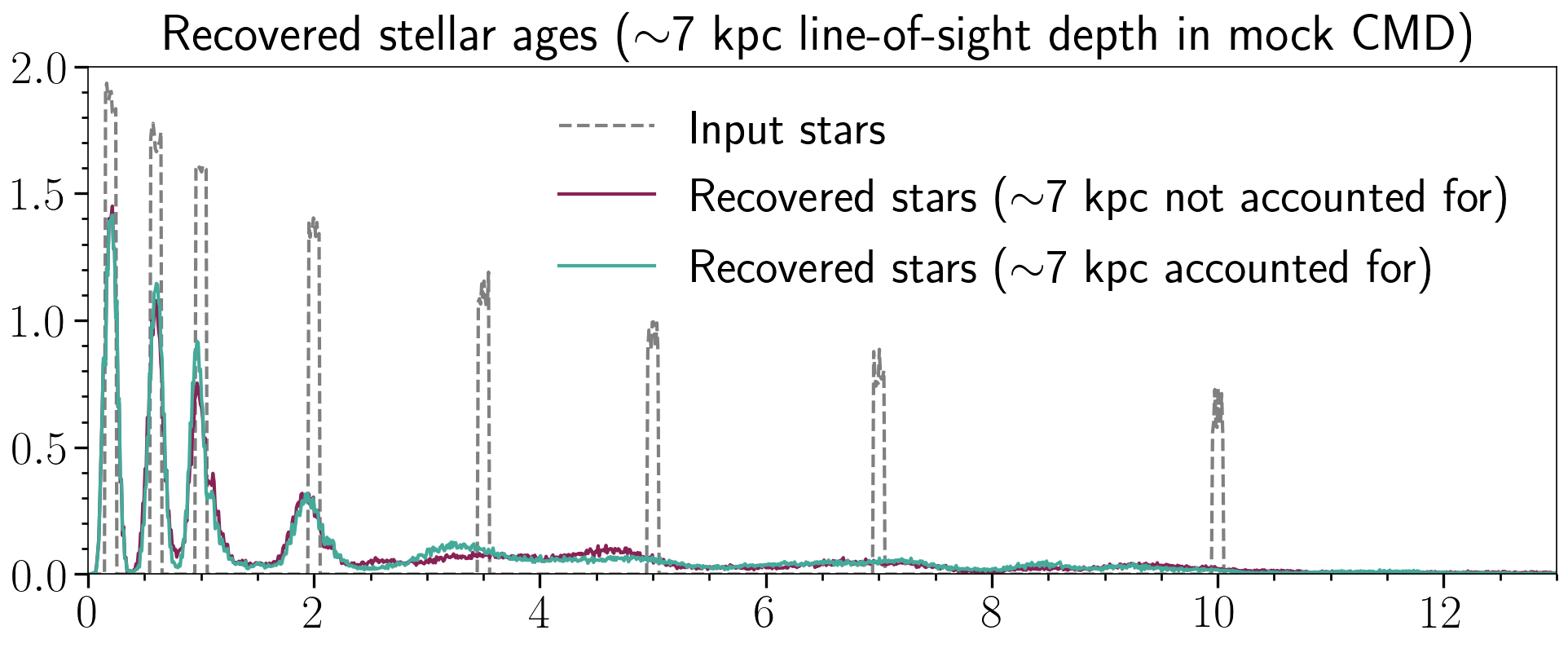}
\includegraphics[scale=0.25]{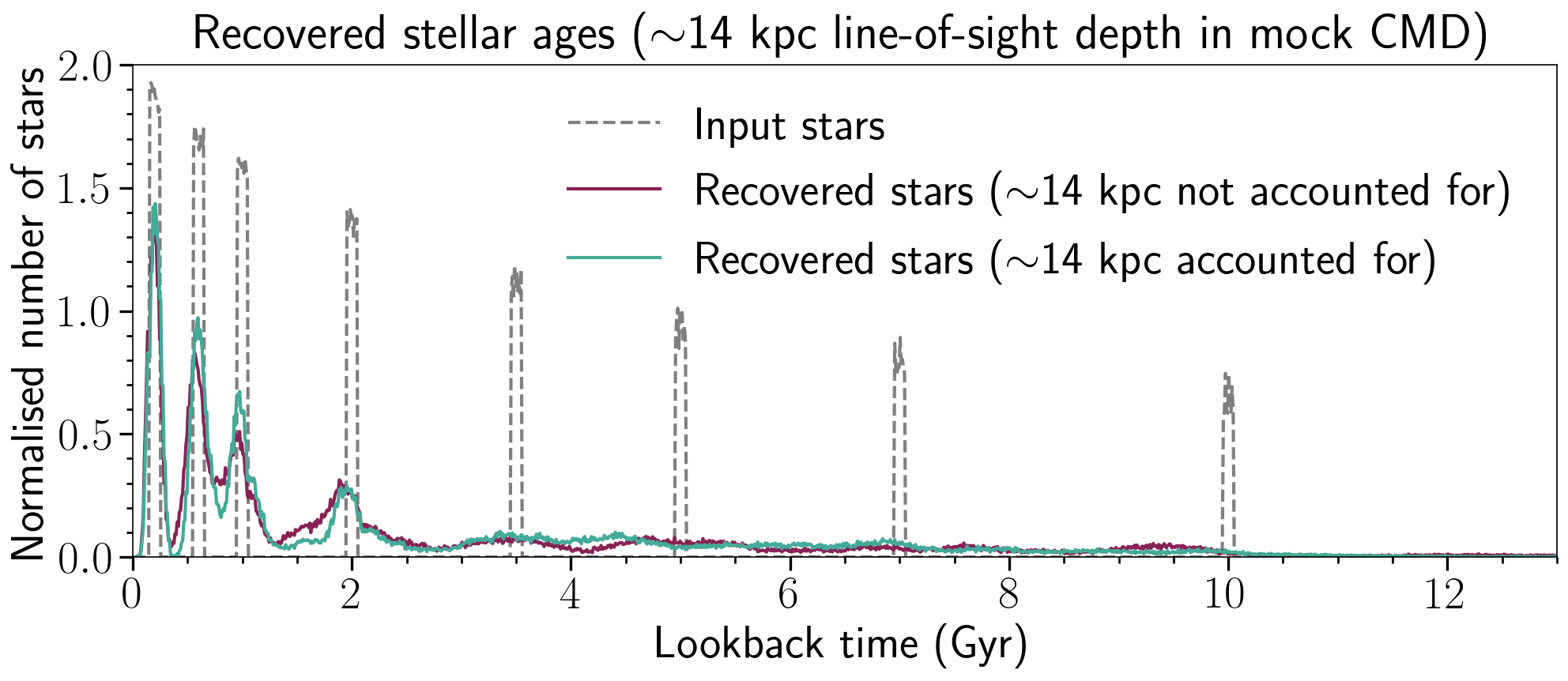}
\includegraphics[scale=0.25]{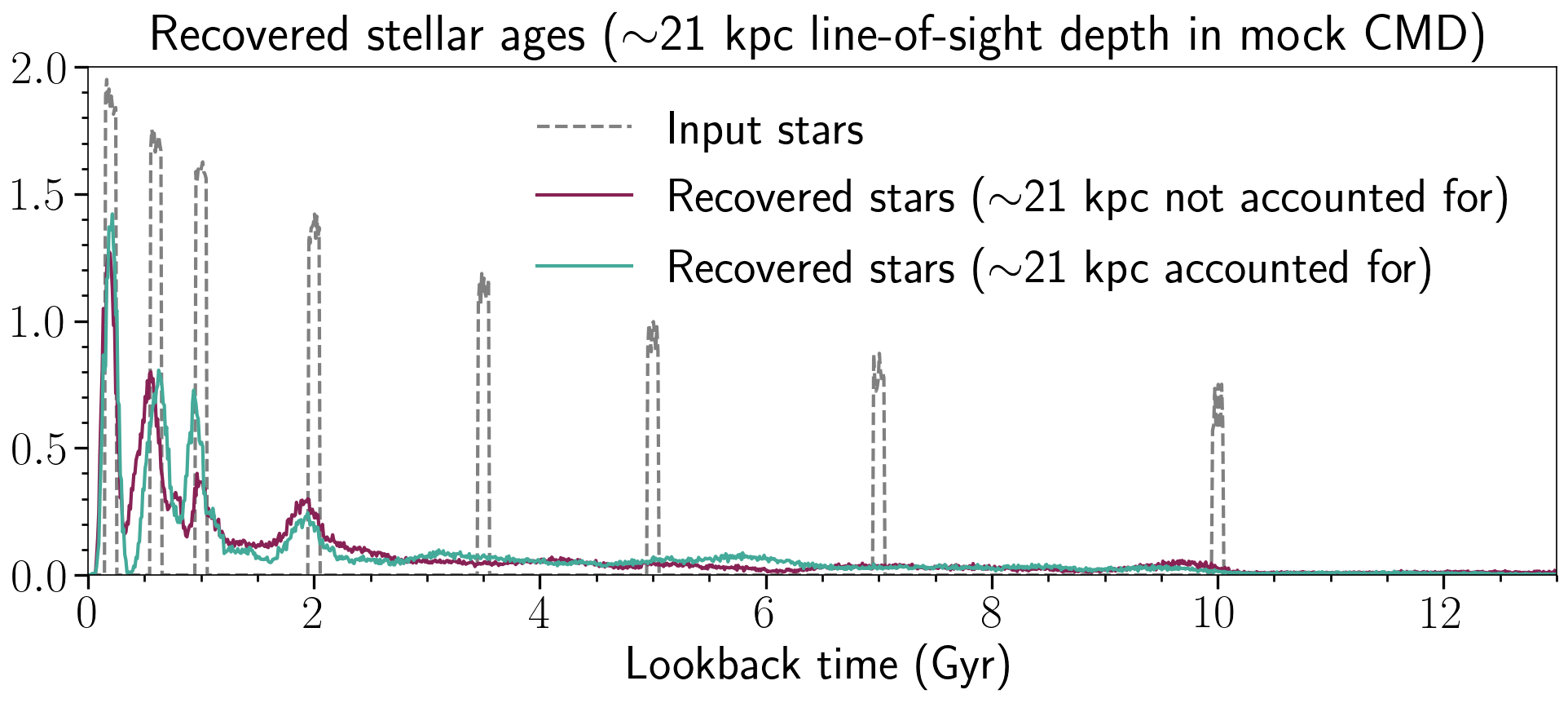}
\caption{Age-resolution tests for a mock CMD with a star formation history consisting of singular bursts. We consider line-of-sight depths of 0, 7, 14 and 21 kpc. In grey are the ages of the stars inputted into THESTORM. In yellow is the result when no line-of-sight depth is present in neither the more nor the synthetic CMD. In purple is the result when a line-of-sight depth is present in the mock CMD and not in the synthetic CMD. In green the line-of-sight is present in mock and synthetic CMD.}\label{fig:FigMocks}
\end{center}
\end{figure*}

\begin{figure*}
\begin{center}
\includegraphics[scale=0.23]{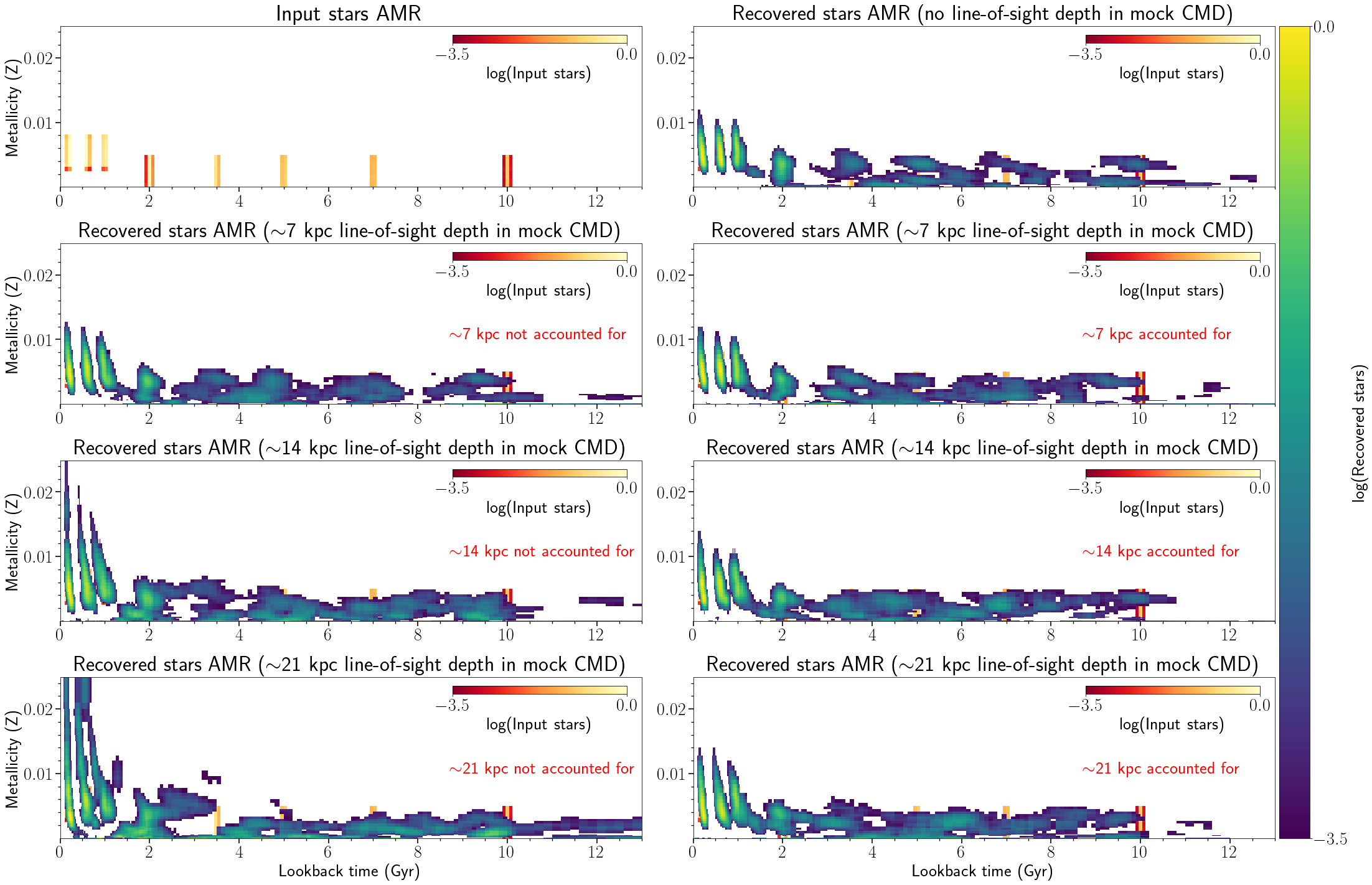}
\caption{Age-metallicity recovery plots for a mock CMD with a star formation history consisting of singular bursts, corresponding to Figure~\ref{fig:FigMocks}. To ease interpretation, we have masked AMR that is less than 3\% of the maximum AMR. First row, left: AMR of input stars mock CMDs for all tests. First, right: AMR recovery when no line-of-sight depth is present in mock CMD and it is not simulated in synthetic CMD, corresponding to top left in Figure \ref{fig:FigMocks}. In grey is the input AMR as reference. Second, third, last row, left: recovered AMR when a line-of-sight depth (7, 14, 21 kpc) is present in the mock and the same line-of-sight depth is \textbf{not} accounted for in the SFH derivation. Second, third, last row, right: recovered AMR when an increasing line-of-sight depth is present in the mock CMD and the same line-of-sight depth \textbf{is} accounted for in the SFH derivation. We see a clear improvement in AMR recovery of younger ages when the line-of-sight depth is accounted for using our methodology. However, as mentioned in text, we believe our resolution decreases in intermediate and old ages and therefore cannot properly test for the effect. We conclude that accounting for the line-of-sight depth becomes more important as the line-of-sight depth increases. To ease interpretation, we have masked AMR that is less than 3\% of the maximum AMR.}\label{fig:FigMocksAMR}
\end{center}
\end{figure*}

\begin{figure*}
\begin{center}
\includegraphics[scale=0.22]{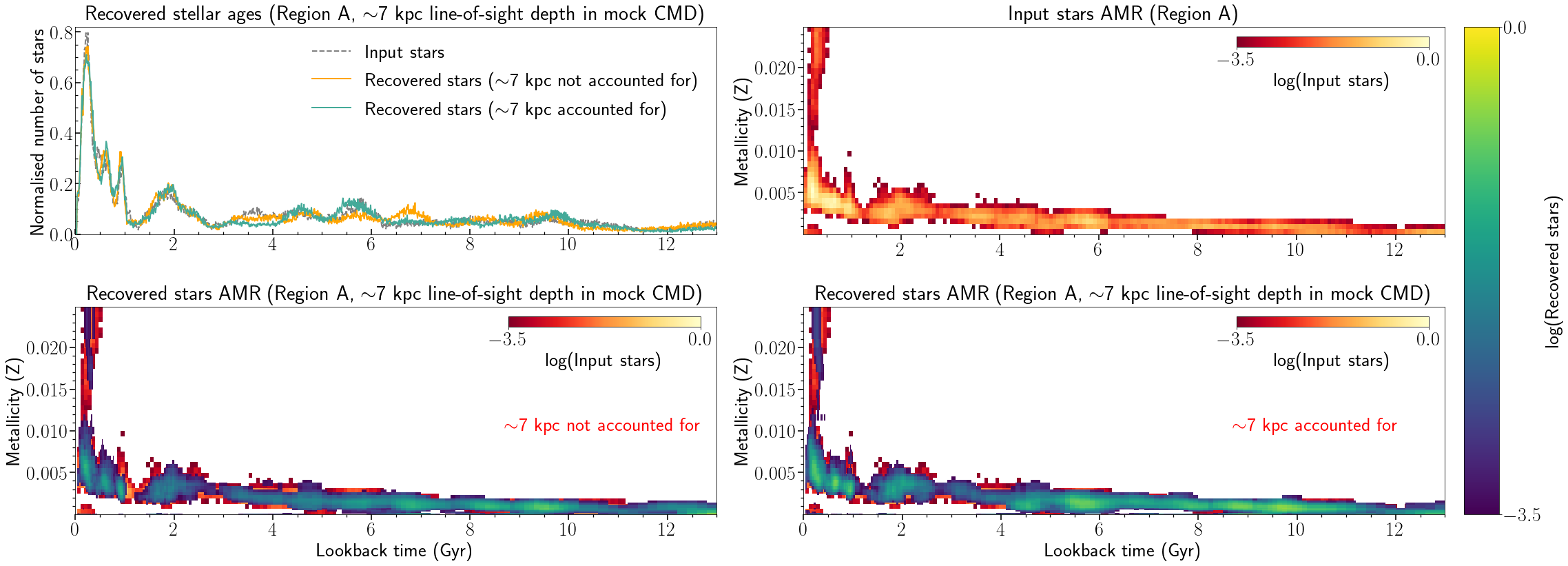}
\caption{Age-resolution tests for a mock CMD with a `realistic' star formation history. The mock CMD used is \hl{region A}'s solution SFH. The solution is post line-of-sight depth simulation, hence the CMD contains a line-of-sight depth of 7 kpc. To ease interpretation, we have masked AMR that is less than 3\% of the maximum AMR. Top, left panel: in grey are the ages of the stars inputted into THESTORM. In yellow is the result when the 7 kpc line-of-sight depth \textbf{is not} accounted for in the fitting process. In green is the result when the line-of-sight depth \textbf{is} accounted for. Top, right: Input AMR of mock CMD. Bottom, left: recovered AMR of mock CMD when line-of-sight depth of 7 kpc \textbf{is not} accounted for. In the background we show the input AMR of the mock CMD (from the top right panel), however, we show it in a red/yellow colour scheme to help the reader distinguish between the input and output stars. Bottom, right: recovered AMR when line-of-sight depth of 7 kpc \textbf{is} accounted for. }\label{fig:FigMocksBursty}
\end{center}
\end{figure*}

\begin{figure*}
\begin{center}
\includegraphics[scale=0.22]{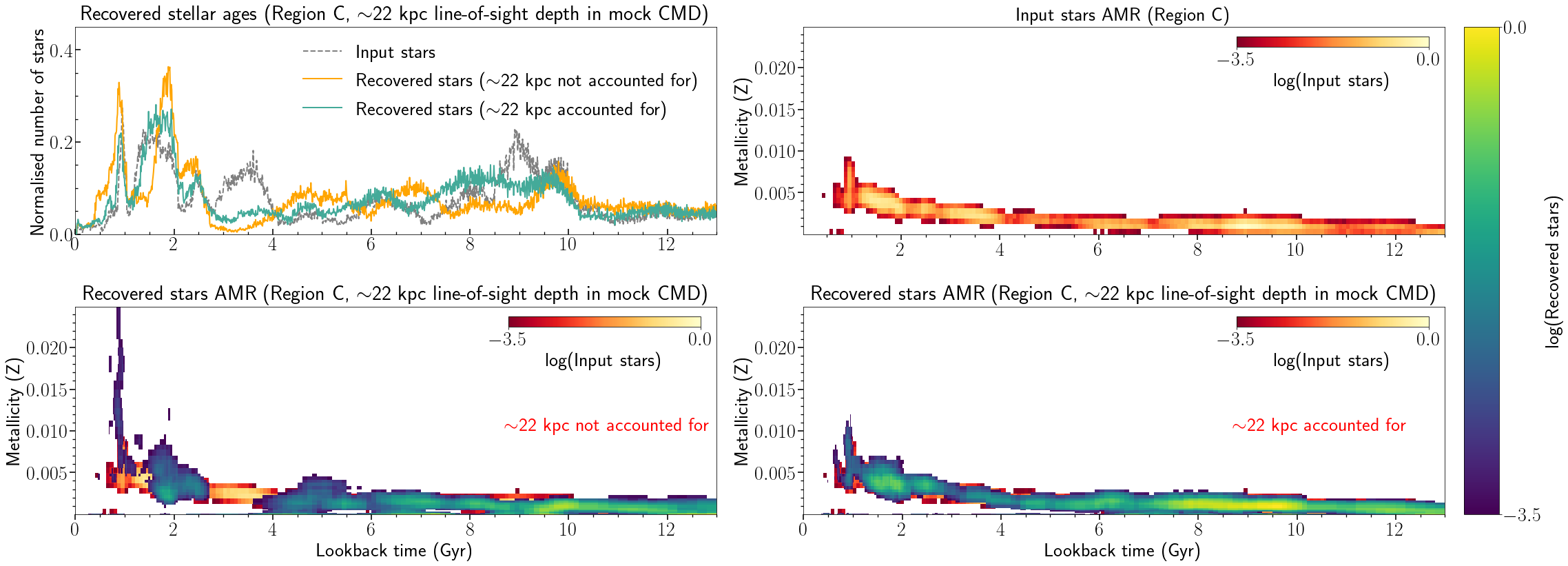}
\caption{\hl{Age-resolution tests for a mock CMD with a `realistic' star formation history and a large line-of-sight depth present within the CMD ($\sim$22 kpc). The mock CMD used is from the SFH results of a region \textit{not} presented in this paper (region C). To ease interpretation, we have masked AMR that is less than 3\% of the maximum AMR. Top, left panel: in grey are the ages of the stars inputted into THESTORM. In yellow is the result when the 22 kpc line-of-sight depth \textbf{is not} accounted for in the fitting process. In green is the result when the line-of-sight depth \textbf{is} accounted for. Top, right: Input AMR of mock CMD. Bottom, left: recovered AMR of mock CMD when line-of-sight depth of 22 kpc \textbf{is not} accounted for. In the background we show the input AMR of the mock CMD (from the top right panel), however, we show it in a red/yellow colour scheme to help the reader distinguish between the input and output stars. Bottom, right: recovered AMR when line-of-sight depth of 22 kpc \textbf{is} accounted for.} }\label{fig:FigMocksBursty22kpc}
\end{center}
\end{figure*}

\subsection{Assessing the effect of the line-of-sight depth on mock data recovery}
\label{subsec:SFHmockslos}
In Section \ref{subsubsec:SFHmockslossingle} we show the first set of tests examining a mock population with star formation bursts at $\sim$0.2 Gyr,  $\sim$0.6 Gyr,  $\sim$1 Gyr,  $\sim$2 Gyr,  $\sim$3.5 Gyr,  $\sim$5 Gyr,  $\sim$7 Gyr, and  $\sim$10 Gyr, each burst 0.1 Gyr wide. For each burst, the metallicity of the stars was set to resemble the range shown in \hl{region A}'s average Z result: Z $=$ 0.003 $-$ 0.008 for the young populations ($\lesssim$1 Gyr), and Z $=$ 0.00001 $-$ 0.005 for the intermediate-age and old populations ($>$ 1 Gyr). We examine CMDs with simulated line-of-sight depths of 0 kpc, 7 kpc, 14 kpc, and 21 kpc; our preliminary results on the line-of-sight depths across the global SMC data do not suggest line-of-sight depths above 23.5 kpc. The line-of-sight simulation follows the procedure described in Section \ref{subsec:sfhsolvelos}. We simulated observational effects applying \verb|DisPar| on these mock CMDs (using the AST results for \hl{region A}), as described in Section \ref{subsec:sfhsolve}. This way we mimic as if our mock CMDs were observed as part of SMASH, making these tests as realistic as possible. The mock CMD was cut in size to match the number of stars in the CMD of \hl{region A} as well (214,527 stars). 

In Section \ref{subsubsec:SFHmockslosbursty}, we show the second set of tests using \hl{region A}'s final SFH output solution CMD as our input mock CMD. The solution CMD already had observational effects simulated in with \verb|DisPar| using \hl{region A}'s line-of-sight depth (7 kpc) as the mother CMD used for the fit had observational errors and the line-of-sight depth simulated.

For both sets of tests, we fit the mock CMD against a synthetic CMD that either contains (case 1) or does not contain (case 2) the same line-of-sight depth as in the input mock CMD. For case 1, both the mock and synthetic CMDs have the same line-of-sight effect simulated, using the method presented in Section \ref{subsec:sfhsolvelos}. This case would mimic our updated method in which line-of-sight is taken into account. For case 2, only the mock CMD has the line-of-sight effect simulated. This case would mimic our original approach, in which the observed data is affected by a distance spread that is not considered in the fitting process.

The fitting procedure closely followed that presented in Section \ref{sec:sfhprocedure}. The same bundle strategy was used (with no bundle 7, given a lack of MW foreground). With these tests we are able to quantify our ability to resolve star formation bursts without the uncertainties caused by foreground stars, distance and reddening guesses, and stellar evolution libraries.

\subsubsection{Recovering single bursts of star formation}
\label{subsubsec:SFHmockslossingle}

This test will allow us to assess our age resolution when computing SFHs as well as testing the effect of line of sight depths. In Figure~\ref{fig:FigMocks} we show the age-resolution tests for a mock CMD with a SFH consisting of bursts which are 100 Myr wide ($\sim$0.2 Gyr,  $\sim$0.6 Gyr,  $\sim$1 Gyr,  $\sim$2 Gyr,  $\sim$3.5 Gyr,  $\sim$5 Gyr,  $\sim$7 Gyr, and  $\sim$10 Gyr). The top left panel in Figure~\ref{fig:FigMocks} shows the recovery of singular bursts (yellow) against the input stellar ages (grey, dashed) with no line-of-sight depth present in the mock nor the synthetic CMD. 
For the case that does not consider depth effects, the recovery of young ages ($\lesssim$1 Gyr) is excellent. The 100 Myr wide peaks are clearly discernible and do not blend with each other. 
After $\sim$1 Gyr we note that the ability to resolve SF bursts degrades for intermediate ages, with bursts difficult to discern beyond 5 Gyr. The reasons for this degradation include photometric errors, a comparatively smaller number of old and intermediate-age populations in the synthetic CMD, and intrinsic uncertainties both due to our method and due to the decreasing age resolution at older ages. Hence, the capacity to resolve short ($\sim$100 Myr) bursts, of the intensity we have simulated in these tests, is limited at intermediate ages.\footnote{For the tests not included in this paper, we show that we are able to recover intense bursts at intermediate and old ages, albeit with a larger dispersion in age. Hence, we are able to draw meaningful, qualitative conclusions from all of the ages shown in this work.}

The top right panel and bottom panels in Figure~\ref{fig:FigMocks} show the case in which the mock CMD was constructed with an added line-of-sight depth, and the synthetic CMD has or has not the line-of-sight depth included (green if added, purple if not added). At 7 kpc, our ability to recover and individually resolve bursts is slightly better when the line-of-sight depth is considered. As the line-of-sight depth increases and becomes more impactful, the improvement when the effect is considered is visible at 14 and 21 kpc with the young peaks blending less with each other. However, given the decreasing resolution, it is not clear how the line-of-sight depth affects the recovery of intermediate-age and old populations.

In Figure \ref{fig:FigMocksAMR} we show the  recovery in the age-metallicity plane (AMR in 2D). In the top left panel we show the input AMR for all of the mocks, and, in the top right panel, we show the AMR recovery when no line-of-sight depth is present in the mock nor in the synthetic CMD. We observe that stars up to 2 Gyr old are recovered very well, and signatures at 3.5 Gyr, 5 Gyr, 7 Gyr, and 10 Gyr old are there. 
The next middle and bottom rows in Figure \ref{fig:FigMocksAMR} examine the effects of the line-of-sight depth  (7 kpc, 14 kpc, and 21 kpc) on the AMR recovery. On the two middle and bottom left panels, we see the recovered AMR when the input mock CMDs have a line-of-sight depth and no line-of-sight depth is accounted for in the synthetic CMD. On the two middle and bottom right panels, the line-of-sight depth is accounted for in the synthetic CMD. It is clear that the inclusion of the line-of-sight depth considerably improves the AMR recovery in all three cases. Accounting for the depth effect is especially significant at large line-of-sight depths: the metallicity estimate of young stars is significantly less spread out and more like the input AMR.

Motivated by these results, in the next section we used \hl{region A}'s final SFH solution CMD as the input mock CMD and examined the recovery. \hl{This way, we test our method and the inclusion of line-of-sight depth on the fit on a more realistic and physically-motivated way.}

\subsubsection{Recovering region A's star formation history}
\label{subsubsec:SFHmockslosbursty}

In Figure \ref{fig:FigMocksBursty} we show the results of the recovery of \hl{region A}'s final SFH. The top left panel shows the age distribution (normalised amount of stars as a function of lookback time), the upper middle panel shows the input AMR, the bottom middle panel shows the AMR not accounting for the line-of-sight depth (7 kpc), and the bottom panel shows the AMR accounting for a 7 kpc line-of-sight depth.
We recover the input ages well up to $\sim$ 3 Gyr (see top panel). In terms of the AMR recovery, both sets of results are also very similar. \hl{On the one hand, we could draw a parallel between this result and the single burst case in Figure \ref{fig:FigMocksAMR} for 7 kpc, and say that both of the results appear within their error because a 7 kpc depth is relatively small. We examine this statement in the next section, where we investigate the SFH recovery of a region affected by a relatively large (22 kpc) line-of-sight depth.}

\subsubsection{Recovering the star formation history of a region affected by a large line-of-sight depth}

\hl{In Figure \ref{fig:FigMocksBursty22kpc} we show the results of the SFH recovery from a region affected by a relatively large (22 kpc) line-of-sight depth. The region (which we call region `C') is also located in the northeastern SMC. From our in prep. results of the spatially-resolved SFH of the SMC using SMASH data (Sakowska et al., in prep.) we do not measure line-of-sight depths larger than 22 kpc in our dataset. We therefore choose this region as representative of the most challenging case we will encounter to examine the robustness of our 2-step SFH procedure against. The top left panel suggests that accounting for the 22 kpc line-of-sight depth (green colour) recovers the input age distribution with less age dispersion. Up to $\sim$3 Gyr, accounting for the effect recovered the widths of the input peaks with more accuracy than when we do not account for it (yellow line). After $\sim$3 Gyr, both results are within their error, albeit the green line does trace the input stellar ages slightly better. In the AMR panels that follow it is clear that accounting for a large line-of-sight depth recovers a more accurate AMR distribution. When the line-of-sight depth is accounted for, the AMR is recovered with less dispersion in metallicity (e.g. at $\sim$1 Gyr) and with less metallicity gaps (e.g. at $\sim$3 Gyr). There is also some tentative evidence that accounting for the line-of-sight depth at intermediate and old ages yields more accuracy- the AMR at $\sim$8 - 10 Gyr is recovered better than when we do not account for the effect. These behaviours are in excellent agreement with the single burst case for a 21 kpc line-of-sight depth in Figure \ref{fig:FigMocksAMR}.}

\hl{We therefore conclude that accounting for the line-of-sight depth during the SFH procedure is worthwhile. Although the regions studied in the paper are affected by line-of-sight depths shown to not significantly change the final SFH (7 kpc), our mock tests show conceptual improvement when the effect is accounted for at increasing line-of-sight depths. Thanks to these findings we have decided to account for the line-of-sight depth in our SFH derivation.}

\subsection{Comparison with SMC line-of-sight depth tests in literature}
\label{subsec:depthlit}
\cite{HarrisZaritsky2004} calculated the SFH on a central 4 $\times$ 4.5$^{\circ}$ area on the SMC's main body using the Magellanic Clouds Photometric Survey (MCPS).  To test for line-of-sight depth effects, they fit a mock CMD with a 12 kpc line-of-sight depth against a synthetic CMD without a line-of-sight depth using $\chi^{2}$ minimisation, concluding that the recovered SFHs were comparable within the errors. 
\cite{Rubele2018} performed a spatially-resolved, SFH determination of a contiguous area of 23.57 deg$^{2}$ of the SMC's main body using near-IR VMC data. They tested for line-of-sight depth effects by fitting the observed CMD against a synthetic CMD with no line-of-sight depth; a synthetic CMD with a line-of-sight depth estimated from \cite{Muraveva2018}; and a synthetic CMD with a best-fitting line-of-sight depth (found by searching a grid of different depth values, which used different reddening and distance values). 
The authors present results for an example region in the northeast (which also contains the SMC's shell). The RR Lyrae distribution in the area equalled $4.3\pm1$ kpc (closely in agreement with our estimate by cross-matching OGLE and SMASH data); the line-of-sight depth in the synthetic CMD was simulated using a Cauchy distribution which considered depths up to 25 kpc.  \cite{Rubele2018} found little to no improvement, with all three solutions within their error bars, noting that estimating the line-of-sight depth using RR Lyrae distribution is limited due to the presence of many different stellar populations within the SMC. 

\vspace{5mm}

\newpage

\bsp	
\label{lastpage}
\end{document}